\documentclass[12pt]{article}

\usepackage{latexsym,amsmath}

\newcounter{subequation}[equation]
\makeatletter

\def\bcite{\@ifnextchar [{\@tempswatrue\@bcitex}{\@tempswafalse\@bcitex[]}}
\def\@bcitex[#1]#2{\if@filesw\immediate\write\@auxout{\string\citation{#2}}\fi
  \let\@bcitea\@empty
  \@bcite{\@for\@bciteb:=#2\do
    {\@bcitea\def\@bcitea{,\penalty\@m\ }%
     \def\@tempa##1##2\@nil{\edef\@bciteb{\if##1\space##2\else##1##2\fi}}%
     \expandafter\@tempa\@bciteb\@nil
     \@ifundefined{b@\@bciteb}{{\reset@font\bf ?}\@warning
       {Citation `\@bciteb' on page \thepage \space undefined}}%
     \hbox{\csname b@\@bciteb\endcsname}}}{#1}}
\def\@bcite#1#2{{#1\if@tempswa , #2\fi}}

\def\thesubequation{\theequation\@alph\c@subequation}
\def\@subeqnnum{{\rm (\thesubequation)}}
\def\slabel#1{\@bsphack\if@filesw {\let\thepage\relax
   \xdef\@gtempa{\write\@auxout{\string
      \newlabel{#1}{{\thesubequation}{\thepage}}}}}\@gtempa
   \if@nobreak \ifvmode\nobreak\fi\fi\fi\@esphack}
\def\subeqnarray{\stepcounter{equation}
\let\@currentlabel=\theequation\global\c@subequation\@ne
\global\@eqnswtrue
\global\@eqcnt\z@\tabskip\@centering\let\\=\@subeqncr
$$\halign to \displaywidth\bgroup\@eqnsel\hskip\@centering
  $\displaystyle\tabskip\z@{##}$&\global\@eqcnt\@ne
  \hskip 2\arraycolsep \hfil${##}$\hfil
  &\global\@eqcnt\tw@ \hskip 2\arraycolsep
  $\displaystyle\tabskip\z@{##}$\hfil
   \tabskip\@centering&\llap{##}\tabskip\z@\cr}
\def\endsubeqnarray{\@@subeqncr\egroup
                     $$\global\@ignoretrue}
\def\@subeqncr{{\ifnum0=`}\fi\@ifstar{\global\@eqpen\@M
    \@ysubeqncr}{\global\@eqpen\interdisplaylinepenalty \@ysubeqncr}}
\def\@ysubeqncr{\@ifnextchar [{\@xsubeqncr}{\@xsubeqncr[\z@]}}
\def\@xsubeqncr[#1]{\ifnum0=`{\fi}\@@subeqncr
   \noalign{\penalty\@eqpen\vskip\jot\vskip #1\relax}}
\def\@@subeqncr{\let\@tempa\relax
    \ifcase\@eqcnt \def\@tempa{& & &}\or \def\@tempa{& &}
      \else \def\@tempa{&}\fi
     \@tempa \if@eqnsw\@subeqnnum\refstepcounter{subequation}\fi
     \global\@eqnswtrue\global\@eqcnt\z@\cr}
\let\@ssubeqncr=\@subeqncr
\@namedef{subeqnarray*}{\def\@subeqncr{\nonumber\@ssubeqncr}\subeqnarray}
\@namedef{endsubeqnarray*}{\global\advance\c@equation\m@ne%
                           \nonumber\endsubeqnarray}

\makeatother

\DeclareFontFamily{OT1}{rsfs10}{}
\DeclareFontShape{OT1}{rsfs10}{m}{n}{ <-> rsfs10 }{}
\DeclareMathAlphabet{\mathscript}{OT1}{rsfs10}{m}{n}

\numberwithin{equation}{section}

\newcommand{\ns}{\normalsize}
\newcommand{\pt}{\partial}

\newcommand{\be}{\begin{equation}}
\newcommand{\ee}{\end{equation}}
\newcommand{\nn}{\nonumber}
\newcommand{\bea}{\begin{eqnarray}}
\newcommand{\eea}{\end{eqnarray}}
\newcommand{\bsea}{\begin{subeqnarray}}
\newcommand{\esea}{\end{subeqnarray}}

\newcommand{\tr}{\textrm{tr}}

\def\a{\alpha}
\def\b{\beta}
\def\g{\gamma}
\def\c{\chi}
\def\d{\delta}
\def\e{\epsilon}

\def\z{\psi}

\def\k{\kappa}
\def\l{\lambda}
\def\m{\mu}
\def\n{\nu}
\def\o{\omega}
\def\p{\pi}

\def\r{\rho}
\def\s{\sigma}

\def\t{\tau}

\def\x{\xi}
\def\z{\zeta}
\def\w{\wedge}

\def\G{\Gamma}

\def\O{\Omega}

\def\cA{{\cal A}}
\def\cF{{\cal F}}
\def\cM{{\cal M}}

\def\brr{\begin{eqnarray}}
\def\err{\end{eqnarray}}
\def\bpl{\Big(}
\def\bpr{\Big)}
\newcommand{\ft}[2]{{\textstyle\frac{#1}{#2}}}

\begin{document}


\begin{titlepage}

\vspace{-2cm}

\title{
   \hfill{\ns UPR-976T} \\[1em]
   {\LARGE Lectures on Heterotic M-Theory}}
\author{
   Burt A.~Ovrut\\[0.5cm]
   {\ns Department of Physics, University of Pennsylvania} \\
   {\ns Philadelphia, PA 19104--6396, USA}}

\maketitle

\begin{abstract}

We present three lectures on heterotic $M$-theory and a fourth lecture
extending this theory to more general orbifolds. In Lecture 1, Ho\v
rava-Witten theory is briefly discussed. We then compactify this theory on
Calabi-Yau threefolds, choosing the ``standard'' embedding of the spin connection
in the gauge connection. We derive, in detail, both the five-dimensional
effective action and the associated actions of the four-dimensional
``end-of-the-world'' branes. Lecture 2 is devoted to showing that this theory
naturally admits static, $N=1$ supersymmetry preserving BPS three-branes, the
minimal vacuum having two such branes. One of these, the ``visible'' brane, is
shown to support a three-generation $E_{6}$ grand unified theory, whereas the
other emerges as the ``hidden'' brane with unbroken $E_{8}$ gauge group.
Thus heterotic $M$-theory emerges as a fundamental paradigm for so-called
``brane world'' scenarios of particle physics. In Lecture 3 , we introduce the
concept of ``non-standard'' embeddings. These are shown to permit a vast
generalization of allowed vacua, leading on the visible brane to new grand
unified theories, such as $SO(10)$ and $SU(5)$, and to the standard model
$SU(3)_{C} \times SU(2)_{L} \times U(1)_{Y}$. It is demonstrated that non-standard
embeddings generically imply the existence of five-branes in the bulk space.
The physical properties of these bulk branes is discussed in detail. Finally,
in Lecture 4 we move beyond Ho\v rava-Witten theory and consider orbifolds
larger than $S^{1}/{\bf Z \rm}_{2}$. For explicitness, we consider $M$-theory
orbifolds on $S^{1}/{\bf Z \rm}_{2} \times T^{4}/{\bf Z \rm}_{2}$, discussing
their anomaly structure in detail and completely determining both the
untwisted and twisted sector spectra.

\end{abstract}

\thispagestyle{empty}

\end{titlepage}


\section{Lecture 1: The Five-Dimensional Effective Theory}


In this first lecture, we introduce our notation and briefly discuss the
theory of the strongly coupled heterotic sting introduced by Ho\v rava and
Witten. In this theory, there is an eleven-dimensional bulk space bounded on
either end of the $x^{11}$-direction by two ``end-of-the-world''
ten-dimensional nine-branes, each supporting an $N=1$, $E_{8}$ supergauge
theory. We then begin our construction of heterotic $M$-theory by
compactifying the Ho\v rava-Witten theory on a Calabi-Yau threefold. This
leads to a five-dimesional bulk space bounded at the ends of the fifth
dimesion by two end-of-the-world four-dimensional three-branes. Assuming, in
this lecture, the ``standard'' embedding of the spin connection into one of
the $E_{8}$ gauge connections we derive, in detail, both the five-dimensional
bulk space effective action and the associated actions of the four-dimensional
boundary branes. We end this lecture by discussing some of the properties of
this effective theory and explicitly giving the $N=2$ supersymmetry
transformations of the bulk space quantum fields.

We begin by briefly reviewing the description of strongly coupled heterotic
string theory as 11-dimensional supergravity with boundaries, as given by
Ho\v{r}ava and Witten~\cite{hw1,hw2}. Our conventions are as follows. We will
consider eleven-dimensional spacetime compactified on a Calabi-Yau space $X$,
with the subsequent reduction down to four dimensions effectively provided by
a double-domain-wall background, corresponding to an $S^1/Z_2$ orbifold. We
use coordinates $x^{I}$ with indices $I,J,K,\cdots = 0,\cdots ,9,11$ to
parameterize the full 11--dimensional space $M_{11}$. Throughout these lectures,
when we refer to orbifolds, we will work in the ``upstairs'' picture
with the orbifold $S^1/Z_2$ in the $x^{11}$--direction. We choose the range
$x^{11}\in [-\pi\rho ,\pi\rho ]$ with the endpoints being identified. The
$Z_2$ orbifold symmetry acts as $x^{11}\rightarrow -x^{11}$. Then there exist
two ten--dimensional hyperplanes fixed under the $Z_2$ symmetry which we
denote by $M_{10}^{(i)}$, $i=1,2$. Locally, they are specified by the
conditions $x^{11}=0,\pi\rho$. Barred indices
$\bar{I},\bar{J},\bar{K},\cdots = 0,\cdots ,9$ are used for the
ten--dimensional space orthogonal to the orbifold.
We use indices $A,B,C,\cdots =4,\cdots 9$ for the Calabi--Yau space.
All fields will be required to have a definite behaviour under the $Z_2$
orbifold symmetry in $D=11$. We demand a bosonic field $\Phi$ to be
even or odd; that is, $\Phi (x^{11})=\pm\Phi (-x^{11})$. For a spinor
$\Psi$ the condition is $\G_{11}\Psi (-x^{11})=\Psi (x^{11})$ so that
the projection to one of the orbifold planes leads to a
ten--dimensional Majorana--Weyl spinor with positive chirality.
Spinors in eleven dimensions will be Majorana spinors with 32 real components
throughout the paper.

The bosonic part of the action is of the form
\begin{equation}
\label{action}
   S = S_{\rm SG}+S_{\rm YM}
\end{equation}
where $S_{\rm SG}$ is the familiar 11--dimensional supergravity
\begin{equation}
 S_{\rm SG} = -\frac{1}{2\k^2}\int_{M^{11}}\sqrt{-g}\left[
                    R+\frac{1}{24}G_{IJKL}G^{IJKL}
           +\frac{\sqrt{2}}{1728}\e^{I_1...I_{11}}
               C_{I_1I_2I_3}G_{I_4...I_7}G_{I_8...I_{11}} \right]
 \label{SSG}
\end{equation}
and $S_{\rm YM}$ are the two $E_8$ Yang--Mills theories on the orbifold planes
explicitly given by
\begin{multline}
   \label{SYM}
   S_{\rm YM} = - \frac{1}{8\pi\k^2}\left(\frac{\k}{4\pi}\right)^{2/3}
        \int_{M_{10}^{(1)}}\sqrt{-g}\;\left\{
           \tr(F^{(1)})^2 - \frac{1}{2}\tr R^2\right\} \\
        - \frac{1}{8\pi\k^2}\left(\frac{\k}{4\pi}\right)^{2/3}
           \int_{M_{10}^{(2)}}\sqrt{-g}\;\left\{
               \tr(F^{(2)})^2 - \frac{1}{2}\tr R^2\right\}\; .
\end{multline}
Here $F_{\bar{I}\bar{J}}^{(i)}$ are the two $E_8$ gauge field strengths and
$C_{IJK}$ is the 3--form with field strength
$G_{IJKL}=24\,\partial_{[I}C_{JKL]}$. In order for the above
theory to be supersymmetric and anomaly free, the Bianchi
identity for $G$ should be modified such that
\begin{equation}
 (dG)_{11\bar{I}\bar{J}\bar{K}\bar{L}} = -4\sqrt{2}\pi
    \left(\frac{\k}{4\pi}\right)^{2/3} \left\{
       J^{(1)}\d (x^{11}) + J^{(2)}\d (x^{11}-\pi\r )
       \right\}_{\bar{I}\bar{J}\bar{K}\bar{L}} \label{Bianchi}
\end{equation}
where the sources are given by
\begin{equation}
 J^{(i)}
    = \frac{1}{16\pi^{2}}\left( {\rm tr}F^{(i)}\wedge F^{(i)}
      - \frac{1}{2}{\rm tr}R\wedge R \right)\; .\label{J}
\end{equation}
Under the $Z_2$ orbifold symmetry, the field components $g_{\bar{I}\bar{J}}$,
$g_{11,11}$, $C_{\bar{I}\bar{J}11}$ are even, while $g_{\bar{I}11}$,
$C_{\bar{I}\bar{J}\bar{K}}$ are odd.

The modification of the right hand side of equation~\eqref{Bianchi} has
important consequences.  While the standard embedding of the
spin connection of the Calabi--Yau threefold into the gauge connection
\begin{equation}
 \tr F^{(1)}\w F^{(1)} = \tr R\w R
\label{condition}
\end{equation}
leads to vanishing source terms in the weakly coupled heterotic string
Bianchi identity (which, in turn, allows one to set the antisymmetric tensor
gauge field to zero), in the present case, one is left with non--zero sources
$\pm\tr R\w R$ on the two hyperplanes. This follows from
the fact that the sources in the Bianchi identity~\eqref{Bianchi} are
located on the orbifold planes with the gravitational part distributed
equally between the two planes. The consequence is that not all components of
the antisymmetric tensor field $G$ can vanish. We find, for the standard
embedding~\eqref{condition}, that all components of $G$ vanish with the
exception of
\begin{equation}
 G_{ABCD} = -\frac{1}{6}\a\,{\e_{ABCD}}^{EF}\,\o_{EF}\,\e (x^{11})
\label{corr}
\end{equation}
where
\begin{equation}
 \a = \frac{1}{8\sqrt{2}\p
v^{2/3}}\left(\frac{\k}{4\p}\right)^{2/3}\int_{{\cal{C}}_{\omega}}
{\tr R\wedge R}.
\label{alpha}
\end{equation}
Here $\e (x^{11})$ is the step function which is $+1$ ($-1$) for $x^{11}$
positive (negative) and
\begin{equation}
 v=\int_X\sqrt{\Omega}
 \label{new1}
\end{equation}
where $\Omega_{AB}$ is a fixed Calabi--Yau metric and $v$ is the associated
volume of the Calabi--Yau threefold. The two--form $\omega_{AB}$ is the K\"ahler
form associated with $\Omega_{AB}$ (that is, $\o_{a\bar{b}}=i\Omega_{a\bar{b}}$
where $a$ and $\bar{b}$ are holomorphic and anti-holomorphic indices) and
${\cal{C}}_{\omega}$ is the Poincare dual four-cycle of $\omega$.
Furthermore, in deriving this result, we have turned off all Calabi--Yau moduli
with the exception of the radial breathing mode. This will be sufficient for
all applications dealing with the universal moduli.

Phenomenologically, there is a regime where the universe appears
five-dimensional. We would, therefore, like to derive an effective
theory in the space consisting of the usual four space-time dimensions
and the orbifold. We will, for simplicity, consider the
universal zero modes only; that is, the five--dimensional graviton
supermultiplet and the breathing mode of the Calabi--Yau space, along
with its superpartners. These form a hypermultiplet in five
dimensions. Furthermore, to keep the discussion as straightforward as possible,
we will not consider boundary gauge matter fields. This simple
framework suffices to illustrate our main ideas and was presented as such
in~\cite{losw1}. The general case was
presented in~\cite{losw2}. Our five-dimensional conventions are the following.
Upon reduction on the Calabi-Yau space we have a five-dimensional spacetime
$M_5$ labeled by indices $\a ,\b ,\g ,\cdots  = 0,\cdots ,3,11$. The
orbifold fixed planes become four-dimensional with indices
$\m,\n,\rho,\cdots = 0,\cdots ,3$.
The 11-dimensional Dirac--matrices $\G^I$ with $\{\G^I,\G^J\}=2g^{IJ}$
are decomposed as $\G^I = \{\g^\a\otimes\l ,{\bf 1}\otimes\l^A\}$
where $\g^\a$ and $\l^A$ are the five-- and six--dimensional Dirac
matrices, respectively. Here, $\l$ is the chiral projection matrix in
six dimensions with $\l^2=1$. In five
dimensions we use symplectic-real spinors~\cite{c0} $\psi^i$ where
$i=1,2$ is an $SU(2)$ index, corresponding to the automorphism
group of the $N=1$ supersymmetry algebra in five dimensions. We will
follow the conventions given in~\cite{GST1}.

We can perform the Kaluza-Klein reduction on the metric
\begin{equation}
 ds_{11}^2 = V^{-2/3}g_{\a\b}dx^\a dx^\b +V^{1/3}\Omega_{AB}dx^Adx^B\; .
 \label{metric1}
\end{equation}
Since the compactification is on a Calabi--Yau manifold, the background
corresponding to metric~\eqref{metric1} preserves eight supercharges, the
appropriate number for a reduction down to
five dimensions. It might appear that we are
simply performing a standard reduction of 11--dimensional supergravity
on a Calabi--Yau space to five dimensions; for example, in the way described in
ref.~\cite{CYred}. There are, however, two important
ingredients that we have not yet included. One is obviously the existence of
the boundary theories. We will return to this point shortly. First, however,
let us explain a somewhat unconventional addition to the bulk theory that
must be included.

Specifically, for the nonvanishing component $G_{ABCD}$ in
eq.~(\ref{corr}) there is no corresponding zero mode
field~\footnote{This can be seen from the mixed part of the Bianchi
identity $\partial_\a G_{ABCD}=0$ which shows that the constant $\a$
in eq.~\eqref{corr} cannot be promoted as stands to a five--dimensional
field. It is possible to dualize in five dimensions so the constant
$\a$ is promoted to a five-form field, but we will not pursue this
formulation here.}.
Therefore, in the reduction, we should take this part of $G$ explicitly
into account. In the terminology of ref.~\cite{gsw}, such an antisymmetric
tensor field configuration is called a ``non--zero mode''. A more recent name
for such a field configuration is a non-vanishing ``G--flux''.
More generally, a non--zero mode is a background antisymmetric tensor field that
solves the equations of motion but, unlike antisymmetric tensor field moduli, has
nonvanishing field strength. Such configurations, for a $p$--form field
strength, can be identified with the cohomology group $H^p(M)$ of the manifold
$M$ and, in particular, exist if this cohomology group is nontrivial. In the
case under consideration, the relevant cohomology group is $H^4(X)$ which
is nontrivial for a Calabi--Yau manifold $X$ since $h^{2,2}=h^{1,1}\geq 1$.
Again, the form of $G_{ABCD}$ in eq.~(\ref{corr}) is somewhat special,
reflecting the fact that we are concentrating here on the universal moduli. In
the general case, $G_{ABCD}$ would be a linear combination of all
harmonic $(2,2)$--forms.

The complete configuration for the antisymmetric tensor field that we use in
the reduction is given by
\bea
 C_{\a\b\g}&,\quad&G_{\a\b\g\d}=24\,\partial_{[\a}C_{\b\g\d]}\nn \\
 C_{\a AB} = \frac{1}{6}\cA_\a\o_{AB}&,\quad&G_{\a\b AB}=\cF_{\a\b}\o_{AB}\; ,
             \quad \cF_{\a\b}=\partial_\a\cA_\b -\partial_\b\cA_\a
              \label{Gmod}\\
 C_{ABC} = \frac{1}{6}\x\o_{ABC}&,\quad&G_{\a ABC}=\partial_\a\x\o_{ABC} \nn
\eea
and the non--zero mode is
\begin{equation}
 G_{ABCD} = -\frac{\a}{6}{\e_{ABCD}}^{EF}\,\o_{EF}\,\epsilon (x^{11})\; ,
            \label{nonzero}
\end{equation}
where $\a$ was defined in eq.~\eqref{alpha}. Here, $\o_{ABC}$ is the
harmonic $(3,0)$ form on the Calabi--Yau space and $\x$ is the corresponding
(complex) scalar zero mode. In addition, we have a five-dimensional vector
field $\cA_\a$ and 3--form $C_{\a\b\g}$, which can be
dualized to a scalar $\s$. The total bulk field content of the
five--dimensional theory is then given by the gravity multiplet
\begin{equation}
(g_{\a\b},\cA_\a ,\psi^i_\a )
\label{burt1}
\end{equation}
together with the universal hypermultiplet
\begin{equation}
(V,\s ,\x ,\bar{\x},\z^i).
\label{burt2}
\end{equation}
Here $\psi_\a^i$ and
$\z^i$ are the gravitini and the hypermultiplet fermions respectively and
$i=1,2$ since they each form a doublet under the $SU(2)$ automorphism group of
$N=2$ supersymmetry in five dimesnions.
>From their relations to the 11--dimensional fields, it is easy to see
that $g_{\m\n}$, $g_{11,11}$, $\cA_{11}$, $\s$ must be even under the
$Z_2$ action whereas $g_{\m 11}$, $\cA_\m$, $\x$ must be odd.

Examples of compactifications with non--zero modes in pure 11--dimensional
supergravity on various manifolds including Calabi--Yau three--folds have
been studied in ref.~\cite{llp}. There is, however, one important way in
which our non--zero mode differs from other non--zero
modes in pure 11--dimensional supergravity. Whereas the latter may be viewed
as an optional feature of generalized Kaluza-Klein reduction, the non--zero
mode in Ho\v{r}ava--Witten theory that we have identified cannot be turned
off. This can be seen from the fact that the constant $\a$ in
expression~\eqref{nonzero} cannot be set to zero, unlike the case in pure
11--dimensional supergravity where it would be arbitrary, since it is fixed by
eq.~\eqref{alpha} in terms of Calabi--Yau data. This fact is, of course,
intimately related to the existence of the boundary source terms, particularly
in the Bianchi identity~\eqref{Bianchi}.

Let us now turn to a discussion of the boundary theories. In the
five--dimensional space $M_5$ of the reduced theory, the orbifold fixed
planes constitute four--dimensional hypersurfaces which we denote by
$M_4^{(i)}$, $i=1,2$. Clearly, since we have used the standard embedding,
there will be an $E_6$ gauge field $A_\m^{(1)}$accompanied by gauginos and
gauge matter fields on the orbifold plane $M_4^{(1)}$. For simplicity,
we will set these gauge matter fields to zero in the following. The field
content of the orbifold plane $M_4^{(2)}$ consists of an $E_8$ gauge field
$A_\m^{(2)}$ and the corresponding gauginos. In addition, there is another
important boundary effect which results from the non--zero internal gauge
field and gravity curvatures. More precisely, for the standard embedding
defined in~\eqref{condition}
\begin{equation}
 \int_X\sqrt{^{6}g}\, \tr F_{AB}^{(1)}F^{(1)AB}
    = \int_X \sqrt{^{6}g}\,\tr R_{AB}R^{AB} = 16\sqrt{2}\p v\left(\frac{4\p}{\k}
                          \right)^{2/3}\a\; ,\qquad
 F_{AB}^{(2)}= 0\; .\label{intcurv}
\end{equation}
In view of the boundary actions~\eqref{SYM}, it follows that we will retain
cosmological type terms with opposite signs on the two boundaries.
Note that the size of those terms is set by the same constant $\a$,
given by eq.~\eqref{alpha}, which determines the magnitude of the non--zero
mode.

We can now compute the five--dimensional effective action of
Ho\v rava--Witten theory. Using the field
configuration~\eqref{metric1}--\eqref{intcurv} we find from the
action~\eqref{action}--\eqref{SYM} that
\begin{equation}
 S_5 = S_{\rm grav}+S_{\rm hyper}+S_{\rm bound}\label{S5}
\end{equation}
where
\bsea
 S_{\rm grav} &=& -\frac{1}{2\k_5^2}\int_{M_5}\sqrt{-g}\left[
                  R+\frac{3}{2}\cF_{\a\b}\cF^{\a\b}+\frac{1}{\sqrt{2}}
                 \e^{\a\b\g\d\e}\cA_\a\cF_{\b\g}\cF_{\d\e}\right] \\
 S_{\rm hyper} &=& -\frac{1}{2\k_5^2}\int_{M_5}\sqrt{-g}\left[
                   \frac{1}{2}V^{-2}\partial_\a V\partial^\a V
                   +2V^{-1}\partial_\a\x\partial^\a\bar{\x}
                   +\frac{1}{24}V^2G_{\a\b\g\d}G^{\a\b\g\d}
                   \right.\nn \\
                && \left.\qquad\qquad\qquad\qquad
                    +\frac{\sqrt{2}}{24}\e^{\a\b\g\d\e}G_{\a\b\g\d}
                   \left(i(\x\partial_\e\bar{\x}-\bar{\x}\partial_\e\x )+
                   2\a\cA_\e\right)+\frac{1}{3}V^{-2}\a^2\right]\qquad\\
 S_{\rm bound} &=& -\frac{1}{2\k_5^2}\left\{2\sqrt{2}\int_{M_4^{(1)}}\sqrt{-g}
                   \, V^{-1}\a-2\sqrt{2}\int_{M_4^{(2)}}\sqrt{-g}\,
                   V^{-1}\a\right\} \nn \\
                && -\frac{1}{16\p\a_{\rm GUT}}
                   \sum_{i=1}^2\int_{M_4^{(i)}}\sqrt{-g}\, V\tr
                   {F_{\m\n}^{(i)}}^2 \; .\label{actparts}
\esea
In this expression, we have now dropped higher-derivative terms. The
4--form field strength $G_{\a\b\g\d}$ is subject to the Bianchi identity
\begin{equation}
 (dG)_{11\m\n\r\s} = -\frac{2\sqrt{2}\pi\k_5^2}{\a_{\rm GUT}}\left\{
       J^{(1)} \d (x^{11})+ J^{(2)} \d (x^{11}-\pi\r )
       \right\}_{\m\n\r\s} \label{Bianchi5}
\end{equation}
which follows directly from the 11--dimensional Bianchi
identity~\eqref{Bianchi}. The currents $J^{(i)}$ have been defined in
eq.~\eqref{J}. The five--dimensional Newton constant $\k_5$ and the
Yang--Mills coupling $\a_{\rm GUT}$ are expressed in terms of
11--dimensional quantities as
\begin{equation}
 \k_5^2=\frac{\k^2}{v}\; ,\qquad \a_{\rm GUT} = \frac{\k^2}{2v}\left(
   \frac{4\p}{\k}\right)^{2/3}\; .
\end{equation}
We have checked the consistency of the truncation which leads
to the above action by an explicit reduction of the 11--dimensional
equations of motion to five dimensions. Note that the potential terms in the
bulk and on the boundaries arise precisely from the inclusion of the
non--zero mode and the gauge and gravity field strengths, respectively.
Since we have compactified on a Calabi--Yau space, we expect
the bulk part of the above action to have eight preserved supercharges
and, therefore, to correspond to minimal $N=1$ supergravity in five
dimensions. Accordingly, let us compare the result \eqref{actparts} to the known
$N=1$ supergravity--matter theories in five
dimensions~\cite{GST1,cn,GST2,Sierra}.

In these theories, the scalar fields in the universal hypermultiplet
parameterize a quaternionic manifold with coset structure
$\cM_Q=SU(2,1)/SU(2)\times U(1)$. Hence, to compare our action to these
we should dualize the three--form $C_{\a\b\g}$ to a scalar field $\s$ by
setting (in the bulk)
\begin{equation}
 G_{\a\b\g\d} = \frac{1}{\sqrt{2}}V^{-2}\e_{\a\b\g\d\e}\left(\partial^\e\s
                -i(\x\partial^\e\bar{\x}-\bar{\x}\partial^\e\x )-2
                \a \epsilon(x^{11})\cA^\e\right)\; .
\end{equation}
Then the hypermultiplet part of the action (\ref{actparts}b) can be written as
\begin{equation}
 S_{\rm hyper} = -\frac{v}{2\k^2}\int_{M_5}\sqrt{-g}\left[ h_{uv}\nabla_\a
                 q^u\nabla^\a q^v +\frac{1}{3}V^{-2}\a^2\right]
 \label{hyper}
\end{equation}
where $q^u=(V,\s ,\x ,\bar{\x})$. The covariant derivative $\nabla_\a$ is
defined as $\nabla_\a q^u= \partial_\a q^u-\a \epsilon(x^{11})\cA_\a k^u$ with
$k^u=(0,-2,0,0)$. The sigma model metric $h_{uv} = \partial_u\partial_vK_Q$
can be computed from the K\"ahler potential
\begin{equation}
 K_Q=-\ln (S+\bar{S}-2C\bar{C})\; ,\quad S=V+\x\bar{\x}+i\s\; ,
     \quad C=\x\; .
\end{equation}
Consequently, the hypermultiplet scalars $q^u$ parameterize a K\"ahler
manifold with metric $h_{uv}$. It can be demonstrated that $k^u$ is a Killing vector
on this manifold. Using the expressions given in ref.~\cite{strom}, one can
show that this manifold is quaternionic with coset structure $\cM_Q$.
Hence, the terms in eq.~\eqref{hyper} that are independent of $\a$ describe the
known form of the universal hypermultiplet action. How do we interpret
the extra terms in the hypermultiplet action depending on
$\a$? A hint is provided by the fact that one of these $\a$-dependent terms
modifies the flat derivative in the kinetic energy to
a generalized derivative $\nabla_\a$. This is exactly the
combination that we would need if one wanted to gauge the $U(1)$ symmetry on
$\cM_Q$ corresponding to the Killing vector $k^u$, using the gauge field
$\cA_\a$ in the gravity supermultiplet. In fact,
investigation of the other terms in the action, including the
fermions, shows that the resulting five-dimensional theory is
precisely a gauged form of supergravity. Not only is a $U(1)$ isometry
of $\cM_Q$ gauged, but at the same time a $U(1)$ subgroup
of the $SU(2)$ automorphism group is also gauged.

What about the remaining $\a$-dependent potential term in the
hypermultiplet action? From $D=4$, $N=2$  theories, we are used to the
idea that gauging a symmetry of the quaternionic manifold describing
hypermultiplets generically introduces potential terms into the
action when supersymmetry is preserved (see for
instance~\cite{andetal}). Such potential terms can be thought of as
the generalization of pure Fayet-Iliopoulos terms. This is precisely what
happens in our theory as well, with the gauging of the  $U(1)$ subgroup
inducing the $\a$-dependent potential term in ~\eqref{hyper}.
The general gauged action was discussed in  detail in~\cite{losw2}.

The phenomenon that the inclusion of non-zero
modes leads to gauged supergravity theories has already been observed in type
II Calabi-Yau compactifications~\cite{sp,michelson}. From the form of the
Killing vector, we see that it is only the scalar
field $\s$, dual to the 4--form $G_{\a\b\g\d}$, which is charged
under the $U(1)$ symmetry. Its charge is fixed by $\a$. We note that
this charge is quantized since, suitably normalized, $\tr R\w R$ is an
element of $H^{2,2}(X,{\bf Z})$.

To analyze the supersymmetry properties of the solutions shortly to be
discussed, we need the supersymmetry variations of the fermions
associated with the theory~\eqref{S5}. They can be obtained either by
a reduction of the 11--dimensional gravitino variation or
by generalizing the known five--dimensional transformations~\cite{GST1,Sierra}
by matching onto gauged four--dimensional $N=2$ theories. It is
sufficient for our purposes to keep the bosonic terms only. Both approaches
lead to
\bea
 \d \psi_\a^i &=& D_\a\e^i
     + \frac{\sqrt{2}i}{8}
          \left({\g_\a}^{\b\g}-4\d_\a^\b\g^\g\right)\cF_{\b\g}\e^i
     - \frac{1}{2}V^{-1/2}\left(
        \partial_\a\x\,{(\t_1-i\t_2)^i}_j
        - \partial_\a{\bar\x}\,{(\t_1+i\t_2)^i}_j \right) \e^j
     \nn \\ &&
     - \frac{\sqrt{2}i}{96}V{\e_\a}^{\b\g\d\e}G_{\b\g\d\e}{(\t_3)^i}_j\e^j
     + \frac{\sqrt{2}}{12}\a V^{-1}\e (x^{11})\g_\a{(\t_3)^i}_j\e^j
     \nn \\
 \d\z^i &=& \frac{\sqrt{2}}{48}V\e^{\a\b\g\d\e}G_{\a\b\g\d}\g_\e\e^i
     - \frac{i}{2}V^{-1/2}\g^\a\left(
        \partial_\a\x\,{(\t_1-i\t_2)^i}_j
        + \partial_\a{\bar\x}\,{(\t_1+i\t_2)^i}_j \right) \e^j
     \label{susy5} \\ &&
     + \frac{i}{2}V^{-1}\g_\b\partial^\b V\e^i
     + \frac{i}{\sqrt{2}}\a V^{-1}\e (x^{11}){(\t_3)^i}_j\e^j \nn
\eea
where $\t_i$ are the Pauli spin matrices.

In summary, we see that the relevant five-dimensional effective theory
for the reduction of Ho\v{r}ava-Witten theory is a gauged $N=1$ supergravity
theory
with bulk and boundary potentials.


\section{Lecture 2: The Domain Wall Solution and Generalizations}


In the second lecture, we show that the effective five-dimensional bulk space
theory does not have flat space for its static vacuum. Instead, the theory
naturally admits static, $N=1$ supersymmetry preserving BPS three-branes,
the minmal vacuum consisting of two end-of-the-world three-branes. One of
these branes, the one with the spin connection embedded in the gauge
connection, supports a three generation $E_{6}$ grand unified theory and,
hence, is called the ``visible'' or physical brane. The other brane is the
``hidden'' brane with an unbroken $E_{8}$ supergauge theory. Thus, heterotic
$M$-theory emerges as a fundamental paradigm for so-called ``brane world''
scenarios of particle physics. In the second part of this lecture, we
generalize the results of Lecture 1 to include, not just the universal
hypermultiplet, but all (1,1)-moduli in the bulk space, as well as matter
scalar multiplets on the boundary three-branes.

In order to re-construct the $D=4$, $N=1$ effective theory originally
discussed in~\cite{w,bd,hor,ssea,choi,hp,aq2,ckm,gcd1,noy,low3,li,mike},
we expect there to be a three--brane
domain wall in five dimensions with a worldvolume lying in the four
uncompactified directions. These solutions should break half
the supersymmetry of the five--dimensional bulk theory and preserve Poincar\'e
invariance in four dimensions.This domain wall can be viewed as the ``vacuum'' of
the five--dimensional theory, in the sense that it provides the appropriate
background for a reduction to the $D=4$, $N=1$ effective theory.

We notice that the theory~\eqref{S5} has all of the prerequisites necessary for
such a three--brane solution to exist. Generally, in order to have a
$(D-2)$--brane in a $D$--dimensional theory, one needs to have a $(D-1)$--form
field or, equivalently, a cosmological constant. This is familiar from the
eight--brane~\cite{8brane} in the massive type IIA supergravity in ten
dimensions~\cite{romans}, and has been systematically studied for theories in
arbitrary dimension obtained by generalized (Scherk-Schwarz) dimensional
reduction~\cite{dom}. In our case, this cosmological term is provided by
the bulk potential term in the action~\eqref{S5}. From the
viewpoint of the bulk theory, we could have multi three--brane solutions with
an arbitrary number of parallel branes located at various places in the
$x^{11}$ direction. As is well known, however, elementary brane solutions have
singularities at the location of the branes, needing to be supported by
source terms. The natural candidates for those source terms, in our case, are
the boundary actions. Given the anomaly-cancellation requirements,
this restricts the possible solutions to those representing a pair of
parallel three--branes corresponding to the orbifold planes.

>From the above discussion, it is clear that in order to find a three-brane
solution, we should start with the Ansatz
\bea
 ds_5^2 &=& a(y)^2dx^\m dx^\n\eta_{\m\n}+b(y)^2dy^2  \\
 V &=& V(y)\nn
\eea
where $a$ and $b$ are functions of $y=x^{11}$ and all other fields vanish.
The general solution for this Ansatz, satisfying the equations of motion
derived from action~\eqref{S5}, is given by
\bea
 a &=&a_0H^{1/2}\nn \\
 b &=& b_0H^2\qquad\qquad H=-\frac{\sqrt{2}}{3}\a|y|+c_0 \label{sol}\\
 V &=&b_0H^3 \nn
\eea
where $a_0$, $b_0$ and $c_0$ are constants. We note that the boundary
source terms have fixed the form of the harmonic function $H$ in the
above solution. Without specific information about the sources, the function
$H$ would generically be glued together from
an arbitrary number of linear pieces with
slopes $\pm\frac{\sqrt{2}}{3}$$\a$. The edges of each piece would then indicate
the location of the source terms. The necessity of matching the boundary
sources at $y=0$ and $\p\r$, however, has forced us to consider only two such
linear pieces, namely $y\in [0,\p\r ]$ and $y\in [-\p\r ,0]$. These pieces are
glued together at $y=0$ and $\p\r$
(recall here that we have identified $\p\r$ and $-\p\r$). Therefore, we
have
\bea
\partial_y^2H &=& -\frac{2\sqrt{2}}{3}\a(\d (y)-\d (y-\p\r ))
\eea
which shows
that the solution represents two parallel three--branes located at the
orbifold planes.
We stress that this solution solves the five--dimensional
theory~\eqref{S5} exactly, and is valid to all orders in $\k$.

Of course, we still have to check that our solution preserves
half of the supersymmetries. When $g_{\a\b}$ and $V$ are the only non--zero
fields, the supersymmetry transformations~\eqref{susy5} simplify to
\bea
 \d\psi_\a^i &=&  D_\a\e^i +\frac{\sqrt{2}}{12}\a\,\e (y)V^{-1}\g_\a\,
                  {(\t_3)^i}_j\e^j\nn \\
 \d\z^i &=& \frac{i}{2}V^{-1}\g_\b\partial^\b V\e^i+\frac{i}{\sqrt{2}}\a\,
            \e (y) V^{-1}\, {(\t_3)^i}_j\e^j\nn\; .
\eea
The Killing spinor equations $\d\psi_\a^i =0$, $\d\z^i=0$ are satisfied
for the solution~\eqref{sol} if we require that the spinor $\e^i$ is
given by
\begin{equation}
 \e^i = H^{1/4}\e^i_0\; ,\quad \g_{11}\e^i_0 = (\t_3)^i_j\e^j_0
\end{equation}
where $\e^i_0$ is a constant symplectic Majorana spinor. This shows that
we have indeed found a BPS solution preserving four of the eight bulk
supercharges.

Let us discuss the meaning of this solution in some detail. First, we notice
that it fits into the general scheme of domain wall solutions in various
dimensions. It is, however, a new
solution to the gauged supergravity action~\eqref{S5} in five dimensions
which has not been constructed previously. In addition, its source terms
are naturally provided by the boundary actions resulting from
Ho\v rava--Witten theory. Most importantly, it constitutes the fundamental
vacuum  solution of a phenomenologically relevant theory. The two parallel
three--branes of the solution, separated by the bulk, are oriented in
the four uncompactified space--time dimensions, and carry the physical
low--energy gauge and matter fields. Therefore, from the
low--energy point of view where the orbifold is not resolved the
three--brane worldvolume is identified with four--dimensional space--time.
In this sense the Universe lives on the worldvolume of a three--brane.

Thus far, we have limited the discussion to the universal hypermultiplet only,
coupled to $N=1$ five--dimensional gauged supergravity. This result can be
extended in a straighforward fashion to include all the $(1,1)$ moduli of the
Calabi--Yau threefold. We will not,
however, explicitly include the $(2,1)$ sector as it is largely unaffected by
the specific structure of Ho\v{r}ava--Witten theory.
We now explain the generalized structure of the zero mode fields used in the
reduction to five dimensions. We begin with the bulk space. Including the zero modes,
the metric is given by
\begin{equation}
 ds^2 = V^{-2/3}g_{\a\b}dx^\a dx^\b +g_{AB}dx^Adx^B
\label{metric2}
\end{equation}
where $g_{AB}$ is the metric of the Calabi--Yau space $X$. Its K\"ahler form
is defined by $\o_{a\bar{b}}=ig_{a\bar{b}}$ and can be expanded
in terms of the harmonic $(1,1)$--forms $\o_{iAB}$, $i=1,\cdots ,h^{1,1}$ as
\begin{equation}
 \o_{AB} = a^i\o_{iAB}\; .
\label{ai_def2}
\end{equation}
The coefficients $a^i=a^i(x^\a )$ are the $(1,1)$ moduli of the Calabi--Yau
space. The Calabi--Yau volume modulus $V=V(x^\a )$ is defined by
\begin{equation}
 V=\frac{1}{v}\int_X\sqrt{^6g}
\label{V_def2}
\end{equation}
where $^6g$ is the determinant of the Calabi--Yau metric $g_{AB}$ and $v$ is
defined in~\eqref{new1}. The modulus
$V$ then measures the Calabi--Yau volume in units of $v$. The factor
$V^{-2/3}$ in eq.~\eqref{metric2} has been chosen such that the metric
$g_{\a\b}$ is the five--dimensional Einstein frame metric. Clearly $V$
is not independent of the $(1,1)$ moduli $a^i$ but it can be expressed as
\begin{equation}
 V = \frac{1}{6}{\cal{K}}(a)\; ,\quad {\cal{K}}(a) = d_{ijk}a^ia^ja^k
\end{equation}
where ${\cal{K}}(a)$ is the K\"ahler potential and $d_{ijk}$ are the Calabi--Yau
intersection numbers.

Let us now turn to the zero modes of the antisymmetric tensor field.
We have the potentials and field strengths,
\bea
 C_{\a\b\g}&,\quad& G_{\a\b\g\d}\nn \\
 C_{\a AB}=\frac{1}{6}\cA_\a^i\o_{iAB}
           &,\quad&G_{\a\b AB}=\cF_{\a\b}^i\o_{iAB} \\
 C_{abc} =\frac{1}{6}\x\o_{abc}&,\quad&G_{\a abc}=X_\a\o_{abc}\nn \; .
 \label{Czero2}
\eea
The five--dimensional fields are therefore an antisymmetric
tensor field $C_{\a\b\g}$ with field strength $G_{\a\b\g\d}$, $h^{1,1}$
vector fields $\cA_\a^i$ with field strengths $\cF_{\a\b}^i$ and a
complex scalar $\x$ with field strength $X_\a$ that arises from the harmonic
$(3,0)$ form denoted by $\o_{abc}$. In the bulk the relations
between those fields and their field strengths are simply
\bea
 G_{\a\b\g\d} &=& 24\,\partial_{[\a}C_{\b\g\d ]} \nn \\
 \cF_{\a\b}^i &=& \partial_\a\cA_\b^i-\partial_\b\cA_\a^i \\
 X_\a &=& \partial_\a\x\nn \; .
\eea
These relations, however, will receive corrections from the boundary
controlled by the 11--dimensional Bianchi identity~\eqref{Bianchi}. We will
derive the associated five--dimensional Bianchi identities later.

Next, we should set up the structure of the boundary fields. The starting
point is the standard embedding of the spin connection in
the first $E_8$ gauge group such that
\begin{equation}
 \tr F^{(1)}\w F^{(1)} = \tr R\w R\; .
\label{condition2}
\end{equation}
As a result, we have an $E_6$ gauge field $A_\a^{(1)}$ with field strength
$F_{\m\n}^{(1)}$ on the first hyperplane and an $E_8$ gauge field $A_\m^{(2)}$
with field strength $F_{\m\n}^{(2)}$ on the second hyperplane. In addition,
there are $h^{1,1}$ gauge matter fields from the $(1,1)$ sector on the first
plane. They are specified by
\begin{equation}
 A_b^{(1)} = \bar{A}_b+{\o_{ib}}^cT_{cp}C^{ip}
\end{equation}
where $\bar{A}_b$ is the (embedded) spin connection. Furthermore,
$p,q,r,\ldots =1,\ldots ,27$ are indices in the fundamental ${\bf 27}$
representation of $E_6$ and $T_{ap}$ are the $({\bf 3},{\bf 27})$
generators of $E_8$ that arise in the decomposition under the subgroup
$SU(3)\times E_6$. Their complex conjugate is denoted by
$T^{ap}$. The $C^{ip}$ are $h^{1,1}$ complex scalars in the
${\bf 27}$ representation of $E_6$. Useful traces for these generators
are $\tr (T_{ap}T^{bq})=\d_a^b\d_p^q$ and
$\tr (T_{ap}T_{bq}T_{cr})=\o_{abc}f_{pqr}$ where $f_{pqr}$ is the totally
symmetric tensor that projects out the singlet in ${\bf 27}^3$.

So far, what we have considered is similar to a reduction of pure
11--dimensional supergravity on a Calabi--Yau space, as for example
performed in ref.~\cite{CYred}, with the addition of gauge and gauge
matter fields on the boundaries. An important difference arises, however,
because the standard embedding~\eqref{condition2}, unlike in the case
of the weakly coupled heterotic string, no longer leads to vanishing
sources in the Bianchi identity~\eqref{Bianchi}. Instead, there is a
net five-brane charge, with opposite sources on each fixed plane,
proportional to $\pm\tr R\w R$. The nontrivial components of
the Bianchi identity~\eqref{Bianchi} are given by
\begin{equation}
 (dG)_{11ABCD} = -\frac{1}{4\sqrt{2}\pi}
    \left(\frac{\k}{4\pi}\right)^{2/3} \left\{
       \d (x^{11}) - \d (x^{11}-\pi\r )\right\} (\tr R\w R)_{ABCD}\; .
 \label{Bianchi12}
\end{equation}
As a result, the components $G_{ABCD}$ of the antisymmetric
tensor field are nonvanishing. We find that
\begin{equation}
G_{ABCD} = -\frac{1}{4V}\a^i\, {\e_{ABCD}}^{EF}\,\o_{iEF}\,\e (x^{11})
\label{nonzero2}
\end{equation}
where
\begin{equation}
\a_i =
 \frac{1}{8\sqrt{2}\pi}\left(\frac{\k}{4\p}\right)^{2/3}\frac{1}{v^{2/3}}
 \int_{{\cal{C}}_{i}}{\tr R\w R }\ .
 \label{alpha_def2}
\end{equation}
Here, the four--cycles ${\cal{C}}_{i}$ are the Poincare duals of the harmonic
$(1,1)$--forms $\omega_{i}$.
The index of the coefficient $\a^i$ in the
second part of the first equation has been raised using the inverse of the metric
\begin{equation}
 G_{ij}(a) = \frac{1}{2V}\int_X{\o_i\w (*\o_j)}
\label{CY_metric2}
\end{equation}
on the $(1,1)$ moduli space. Note that, while the coefficients $\a_i$ with
lowered index are truly constants, as is apparent from
eq.~\eqref{alpha_def2}, the coefficients $\a^i$ depend on the $(1,1)$
moduli $a^i$ since the metric~\eqref{CY_metric2} does. We can derive an
expression for the boundary $\tr F^2$ and $\tr R^2$ terms in the action
essential for the reduction of the boundary theories. We have
\begin{equation}
 \tr R_{AB}R^{AB} = \tr F^{(1)}_{AB}F^{(1)AB} =
  4\sqrt{2}\p\left(\frac{4\p}{\k}\right)^{2/3}V^{-1}\a^i\o^{AB}\o_{iAB}
 \label{R22}
\end{equation}
while, of course
\begin{equation}
 \tr F^{(2)}_{AB}F^{(2)AB} = 0\; .
\end{equation}

The expression~\eqref{nonzero2} for $G_{ABCD}$ with $\a_i$ as defined in
~\eqref{alpha_def2} is, as previously discussed, the new and somewhat
unconventional ingredient in our
reduction. This configuration
for the antisymmetric tensor field strength is the generalized nonzero mode or
$G$--flux. Generally,
a nonzero mode is defined as a nonzero internal antisymmetric tensor
field strength $G$ that solves the equation of motion. In contrast,
conventional zero modes of an antisymmetric tensor field, like those
in eq.~\eqref{Czero2}, have vanishing field strength once the moduli fields
are set to constants. Since the kinetic term $G^2$ is positive for a nonzero
mode it corresponds to a nonzero energy configuration. Given that
nonzero modes, for a $p$--form field strength, satisfy
\begin{equation}
 dG=d^*G=0
\end{equation}
they correspond to harmonic forms of degree $p$. Hence, they can
be identified with the $p$th cohomology group $H^p(X)$ of the internal
manifold $X$. In the present case, we are dealing with a four--form
field strength on a Calabi--Yau threefold $X$ so that the relevant
cohomology group is $H^4(X)$. The expression~\eqref{nonzero2} is just an
expansion of the nonzero mode in terms of the basis of
$H^4(X)$. The appearance of all harmonic $(2,2)$ forms shows that it is
necessary to include the complete $(1,1)$ sector into the low energy effective
action in order to fully describe the nonzero mode.
On the other hand, harmonic $(2,1)$ forms do not
appear here and are, hence, less important in our context. We stress that
the nonzero mode~\eqref{nonzero2}, for a given Calabi--Yau space, specifies
a fixed element in $H^4(X)$ since the coefficients $\a_i$ are fixed in terms
of Calabi--Yau properties. Thus we see that, correctly normalized, $G$
is in the integer cohomology of the Calabi-Yau manifold.
We emphasize that in heterotic $M$-theory, we are not free to turn off the
non--zero mode. Its presence is
simply dictated by the nonvanishing boundary sources.

Let us now summarize the field content which we have obtained above and
discuss how it fits into the multiplets of five--dimensional $N=1$
supergravity. We know that the gravitational multiplet should contain
one vector field, the graviphoton. Thus, since the reduction leads to
$h^{1,1}$ vectors, we must have $h^{1,1}-1$ vector multiplets. This
leaves us with the $h^{1,1}$ scalars $a^i$, the complex scalar $\xi$
and the three-form $C_{\a\b\g}$. Since there is one scalar in each
vector multiplet, we are left with three unaccounted for real scalars
(one from the set of $a^i$, and $\xi$) and the three-form. Together,
these fields form the ``universal hypermultiplet;'' universal because
it is present independently of the particular form of the
Calabi-Yau manifold. From this, it is clear that it must be the overall
volume breathing mode  $V=\frac{1}{6}d_{ijk}a^ia^ja^k$ that is the
additional scalar from the set of the $a^i$ which enters the universal
multiplet. The three-form may appear a little unusual, but
recall that in five dimensions a three-form is dual to a scalar
$\s$. Thus, the bosonic sector of the universal hypermultiplet consists
of the four scalars $(V,\s,\xi,\bar{\xi})$, as presented previously.

The $h^{1,1}-1$ vector multiplet scalars are the remaining $a^i$. More
properly, since the breathing mode $V$ is already part of a
hypermultiplet it should be first scaled out when defining the shape
moduli
\begin{equation}
 b^i=V^{-1/3}a^i\; .
\label{bi_def2}
\end{equation}
Note that the $h^{1,1}$ moduli $b^i$ represent only $h^{1,1}-1$
independent degrees of freedom as they satisfy the constraint
\begin{equation}
 {\cal{K}}(b)\equiv d_{ijk}b^ib^jb^k =6\; .
\end{equation}
The graviton and graviphoton of the gravity multiplet are given by
\begin{equation}
(g_{\a\b},\frac{2}{3}b_i\cA^i_\a).
\end{equation}
Therefore, in total, the five dimensional bulk theory contains a gravity
multiplet, the universal hypermultiplet and $h^{1,1}-1$ vector multiplets.
The inclusion of the $(2,1)$ sector of the Calabi--Yau space would lead
to an additional $h^{2,1}$ set of hypermultiplets in the theory. Since
they will not play a prominent r\^ole in our context they will not be
explicitly included in the following.

On the boundary $M^{(1)}_4$ we have an $E_6$ gauge multiplet
$(A_\m^{(1)},\c^{(1)})$ and $h^{1,1}$ chiral multiplets $(C^{ip},\eta^{ip})$
in the fundamental ${\bf 27}$ representation of $E_6$. Here $C^{ip}$ denote
the complex scalars and $\eta^{ip}$ the chiral fermions. The other boundary,
$M^{(2)}_4$, carries an $E_8$ gauge multiplet $(A_\m^{(2)},\c^{(2)})$ only.
Inclusion of the $(2,1)$ sector would add $h^{2,1}$ chiral multiplets in
the $\bf{\overline{\bf 27}}$ representation of $E_6$ to
the field content of the boundary $M_4^{(1)}$. Any even bulk field
will also survive on the boundary. Thus, in addition to the
four--dimensional part of the metric, the scalars $b^i$ together with
$\cA^i_{11}$, and $V$ and $\s$ survive on the boundaries. These pair
into $h^{1,1}$ chiral muliplets.

We are now ready to derive the bosonic part of the
five--dimensional effective action for the $(1,1)$ sector. Inserting the
expressions for the various fields into the 11-dimensional supergravity
action~\eqref{action} and dropping higher derivative terms we find
\begin{equation}
 S_5 = S_{\rm grav,vec}+S_{\rm hyper}+S_{\rm bound}+S_{\rm matter}
 \label{S52}
\end{equation}
with
\bsea
 S_{\rm grav,vec} &=& -\frac{1}{2\k_5^2}\int_{M_5}\sqrt{-g}\left[R+
                      G_{ij}\pt_\a b^i \pt^\a b^j +
                      \right.\nn \\
                   && \qquad\qquad\qquad\qquad \left.
                     G_{ij}\cF_{\a\b}^i\cF^{j\a\b}+\frac{\sqrt{2}}{12}
                      \e^{\a\b\g\d\e}d_{ijk}\cA_\a^i\cF_{\b\g}^j\cF_{\d\e}^k
                      \right]\\
 S_{\rm hyper} &=& -\frac{1}{2\k_5^2}\int_{M_5}\sqrt{-g}\left[
                   \frac{1}{2}V^{-2}\partial_\a V\partial^\a V
                   +2V^{-1}X_\a\bar{X}^\a
                   +\frac{1}{24}V^2G_{\a\b\g\d}G^{\a\b\g\d}
                   \right.\nn \\
                && \qquad\qquad\qquad\qquad
                    +\frac{\sqrt{2}}{24}\e^{\a\b\g\d\e}G_{\a\b\g\d}
                   \left(i(\x\bar{X}_\e-\bar{\x}X_\e )-
                   2\e (x^{11})\a_i\cA_\e^i\right)\nn\\
                &&\left.\qquad\qquad\qquad\qquad
                  +\frac{1}{2}V^{-2}G^{ij}\a_i\a_j\right]\qquad\\
 S_{\rm bound} &=& -\frac{\sqrt{2}}{\k_5^2}\int_{M_4^{(1)}}\sqrt{-g}
                   \, V^{-1}\a_ib^i
                   +\frac{\sqrt{2}}{\k_5^2}\int_{M_4^{(2)}}\sqrt{-g}\,
                   V^{-1}\a_ib^i \\
 S_{\rm matter} &=& -\frac{1}{16\p\a_{\rm GUT}}
                   \sum_{n=1}^2\int_{M_4^{(n)}}\sqrt{-g}\, V\tr
                   {F_{\m\n}^{(n)}}^2\nn \\
                 && -\frac{1}{2\p\a_{\rm GUT}}
                    \int_{M_4^{(1)}}\sqrt{-g}\left[ G_{ij}(D_\m C)^i
                    (D^\m\bar{C})^j\right.\nn\\
                 &&\left.\qquad\qquad\qquad\qquad\qquad\qquad
                    +V^{-1}G^{ij}\frac{\partial W}
                    {\partial C^{ip}}\frac{\partial\bar{W}}
                    {\partial\bar{C}^j_p}+D^{(u)}D^{(u)}\right] \; .
                    \label{actparts2}
\esea
All fields in this action that originate from the 11--dimensional
antisymmetric tensor field are subject to a nontrivial Bianchi identity.
Specifically, from eq.~\eqref{Bianchi} we have
\bsea
 (dG)_{11\m\n\r\s} &=& -\frac{2\sqrt{2}\pi\k_5^2}{\a_{\rm GUT}}\left\{
       J^{(1)} \d (x^{11})+ J^{(2)} \d (x^{11}-\pi\r )
       \right\}_{\m\n\r\s} \\
 (d\cF^i)_{11\m\n} &=& -\frac{\k_5^2}{4\sqrt{2}\pi\a_{\rm GUT}}J_{\m\n}^i
                       \d (x^{11}) \\
 (dX)_{11\m} &=& -\frac{\k_5^2}{4\sqrt{2}\pi\a_{\rm GUT}}J_\m
                 \d (x^{11}) \label{Bianchi52}
\esea
with the currents defined by
\bsea
 J_{\m\n\r\s}^{(n)}  &=& \frac{1}{16 \pi^{2}}\left(\tr F^{(n)}\w F^{(n)}
                        -\frac{1}{2}\tr R\w R
\right)_{\m\n\r\s} \\
 J_{\m\n}^i &=& -2iV^{-1}\G^i_{jk}\left((D_\m C)^{jp}(D_\n\bar{C})^k_p
                -(D_\m\bar{C})^k_p(D_\n C)^{jp}\right) \\
 J_\m &=& -\frac{i}{2}V^{-1}d_{ijk}f_{pqr}(D_\m C)^{ip}C^{jq}C^{kr}\; .
  \label{currents5}
\esea
The five--dimensional Newton constant $\k_5$ and the
Yang--Mills coupling $\a_{\rm GUT}$ are expressed in terms of
11--dimensional quantities as
\begin{equation}
 \k_5^2=\frac{\k^2}{v}\; ,\qquad \a_{\rm GUT} = \frac{\k^2}{2v}\left(
   \frac{4\p}{\k}\right)^{2/3}\; .
\label{fdconst2}
\end{equation}
We still need to define various quantities in the above action. The
metric $G_{ij}$ is given in terms of the K\"ahler potential ${\cal{K}}$ as
\begin{equation}
 G_{ij}=-\frac{1}{2}\frac{\partial}{\partial b^i}\frac{\partial}
           {\partial b^j}\ln {\cal{K}}\; .
\end{equation}
The corresponding connection $\G_{jk}^i$ is defined as
\begin{equation}
 \G_{jk}^i=\frac{1}{2}G^{il}\frac{\partial G_{jk}}{\partial b^l}\; .
\end{equation}
We recall that
\begin{equation}
 {\cal{K}} = d_{ijk}b^ib^jb^k\; ,
\label{kpot2}
\end{equation}
where $d_{ijk}$ are the Calabi--Yau intersection numbers. All indices
$i,j,k,\cdots$ in the five--dimensional theory are raised and lowered with
the metric $G_{ij}$. We recall that the fields $b^i$ are subject to the constraint
\begin{equation}
 {\cal{K}}= 6
\label{b_cons2}
\end{equation}
which should be taken into account when equations of motion are derived
from the above action. Most conveniently, it can be implemented by adding
a Lagrange multiplier term $\sqrt{-g}\l({\cal{K}}(b)-6)$ to the bulk action.
Furthermore, we need to define the superpotential
\begin{equation}
 W=\frac{1}{6}d_{ijk}f_{pqr}C^{ip}C^{jq}C^{kr}
\label{superpot2}
\end{equation}
and the D--term
\begin{equation}
 D^{(u)} =G_{ij}\bar{C}^jT^{(u)}C^i
\label{Dterm2}
\end{equation}
where $T^{(u)}$, $u=1,\ldots ,78$ are the $E_6$ generators in the fundamental
representation. The consistency of the above theory has been explicitly
checked by a reduction of the 11--dimensional equations of motion.

The most notable features of this action, at first sight, are the bulk
and boundary potentials for the $(1,1)$ moduli $V$ and $b^i$ that
appear in $S_{\rm hyper}$ and $S_{\rm bound}$. Those potentials involve
the five--brane charges $\a_i$, defined by eq.~\eqref{alpha_def2}, that
characterize the nonzero mode. The bulk potential in the hypermultiplet
part of the action arises directly from the kinetic term $G^2$ of the
antisymmetric tensor field with the expression~\eqref{nonzero2} for the
nonzero mode inserted. It can therefore be interpreted as the energy
contribution of the nonzero mode. The origin of the boundary potentials,
on the other hand, can be directly seen from eq.~\eqref{R22} and
the 10-dimensional boundary actions. Essentially, they arise because the
standard embedding leads to nonvanishing internal boundary actions due
to the crucial factor $1/2$ in front of the $\tr R^2$ terms. This is in
complete analogy with the appearance of nonvanishing sources in the
internal part of the Bianchi identity which led us to introduce the
nonzero mode. The action presented in~\eqref{S52} and~\eqref{actparts2} was
first derived in~\cite{losw2}.


\section{Lecture 3: Bulk Five-Branes and Non-Standard Embeddings}


In this third lecture, we begin by discussing the simplest BPS three-brane
solution of the generalized five-dimensional heterotic $M$-theory presented in
Lecture 2. We then commence a major extension of heterotic $M$-theory. Until
now, we have employed the standard embedding of the spin connection of the
Calabi-Yau threefold into the gauge connection of the visible brane. However,
unlike the case of the weakly coupled heterotic string, there is nothing
compelling about the standard embedding in heterotic $M$-theory. Quite the
contrary, it is more natural to consider ``non-standard'' embeddings. Here, we
will only briefly discuss such embeddings, referring the reader to the TASI
2001 lectures by Daniel Waldram for details. In this lecture, we focus on
one of the important phenomena assoaciated with non-standard embeddings,
namely, the appearance of one or more bulk space three-branes (actually,
$M$5-branes wrapped on holomorphic curves in the Calabi-Yau threefold). In the
second part of this lecture, we will discuss the existence and properties of
bulk space wrapped five-branes in detail.

We would now like to find the simplest BPS domain wall
solutions of the generalized five--dimensional heterotic $M$-theory.
From the above results, it is clear that the proper Ansatz for
the type of solutions we are looking for is given by
\bea
 ds_5^2 &=& a(y)^2dx^\m dx^\n\eta_{\m\n}+b(y)^2dy^2 \nn \\
 V &=& V(y) \label{3_ans33}\\
 b^i &=& b^i(y)\nn\; ,
\eea
where we use $y=x^{11}$ from now on. A solution to the generalized equations
of motion is somewhat hard to find, essentially due to the complication
caused by the
inclusion of all $(1,1)$ moduli and the associated K\"ahler structure. The
trick is to express the solution in terms of certain functions $f^i=f^i(y)$
which are only implicitly defined rather than trying to find fully explicit
formulae. It turns out that those functions are fixed by the equations
\begin{equation}
 d_{ijk}f^jf^k=H_i\; ,\quad H_i=-2\sqrt{2}k\a_i|y|+k_i\label{beta_def3}
\end{equation}
where $k$ and $k_i$ are arbitrary constants. Then the solution can be written
as
\bea
 V &=&\left(\frac{1}{6}d_{ijk}f^if^jf^k\right)^2 \nn \\
 a &=&\tilde{k}V^{1/6}\nn\\
 b &=& kV^{2/3} \\
 b^i &=&V^{-1/6}f^i\nn
 \label{solution3}
\eea
where $\tilde{k}$ is another arbitrary constant. We have checked that
this solution is indeed a BPS state of the theory; that is, that it
preserves four of the eight supercharges. Note that we have chosen the above
solution to have no singularities other than those
at the two boundaries. Specifically, the
harmonic functions $H_i$ in eq.~\eqref{beta_def3} satisfy
\begin{equation}
 {H_i}'' = 4\sqrt{2}k\a_i(\d (y)-\d (y-\p\r ))\; ,
\end{equation}
indicating sources at the orbifold planes $y=0,\p\r$.
Recall that we have restricted the range of $y$ to $y\in [-\p\r ,\p\r ]$
with the endpoints identified. This explains the second delta--function
at $y=\p\r$ in the above equation.
We conclude that the solution~\eqref{solution3} represents a multi--charged
double domain wall (three--brane) solution with the two walls located
at the orbifold planes. It preserves four--dimensional Poincar\'e invariance
as well as four of the eight supercharges.

In Lecture 2, we have presented a related three--brane solution which
was less general in that it involved the universal Calabi--Yau
modulus $V$ only. Clearly, we should be able to recover this solution
from eq.~\eqref{solution3} if we consider the specific case $h^{1,1}=1$.
Then we have $d_{111}=6$ and it follows from eq.~\eqref{beta_def3} that
\begin{equation}
 f^1 = \left(\frac{\sqrt{2}}{3}k\a_1|y|+k_1\right)^{1/2}\; .
\end{equation}
Inserting this into eq.~\eqref{solution3} provides us with the explicit
solution in this case which is given by
\bea
 a &=& a_0H^{1/2} \nn \\
 b &=& b_0H^2\qquad\qquad H=-\frac{\sqrt{2}}{3}\a |y|+c_0\; ,\quad \a =\a^1
 \label{u_sol3}\\
 V &=& b_0H^3\nn \; .
\eea
The constant $a_0$, $b_0$ and $c_0$ are related to the integration constants
in eq.~\eqref{solution3} by
\begin{equation}
 a_0=\tilde{k}k^{1/2}\; ,\qquad b_0=k^3\; ,\qquad c_0=\frac{k_1}{k}\; .
\end{equation}
Eq.~\eqref{u_sol3} is indeed exactly the solution that was found in
Lecture 2. It still represents a double domain wall. However, in
contrast to the general solution it couples to one charge $\a =\a^1$ only.
Geometrically, it describes a variation of the five--dimensional metric and
the Calabi--Yau volume across the orbifold.

At this point, we introduce an important generalization which greatly
expands the scope, theoretical interest and phenomenological implications of
heterotic $M$-theory. First, note that all of our previous results have assumed
that the gauge field vacuum on the Calabi--Yau threefold is
identical to the geometrical spin connection. That is, we have assumed the
standard embedding defined in~\eqref{condition}. Since any Calabi--Yau
threefold has holonomy group $SU(3)$, it follows that the spin connection and,
hence, the gauge connection has structure group $G=SU(3)$. The
four-dimensional low energy theory then exhibits a gauge group $H$ which
is the commutant of $G$ in $E_{8}$. Since $G=SU(3)$, it follows that
$H=E_{6}$, as we discussed above. Although the choice of the
standard embedding was natural within the context of weakly coupled heterotic
superstring theory, there is no reason to single it out from other gauge vacua in
$M$-theory. Indeed, the only constraint on the gauge vacua in
heterotic $M$-theory is that they be compatible with $N=1$ supersymmetry on
the boundary planes. That is, the gauge connection on the Calabi--Yau
threefold must satisfy the Hermitian Yang--Mills equation, but is otherwise
arbitrary. Clearly, it would be of significant interest to demonstrate the
existence of gauge vacua other than the standard embedding. For example, if
one could construct a ``non--standard'' embedding gauge vacuum with structure
group, say, $G=SU(5) \times {\bf Z \rm}_{2}$, then the low energy gauge group
in four-dimensions would be the standard model group
$SU(3)_{C} \times SU(2)_{L} \times U(1)_{Y}$. Since the Calabi--Yau threefold has a
Euclidean signature and is compact, we will refer to any gauge configuration
with structure group $G \subset E_{8}$ that satisfies the Hermitian
Yang--Mills equation as a $G$-instanton. We will, therefore, expand the vacua
of heterotic $M$-theory by compactifying Ho\v rava-Witten theory on
Calabi--Yau manifolds with $G$-instantons.

Initially, this seems to be a very difficult task, since not a single solution
to the Hermitian Yang--Mills equations on a Calabi--Yau threefold is known,
with the exception of the standard embedding. However, at this point, some
important mathemetical results become relevant, which allow us to demonstrate
the existence and compute the properties of very large classes of
$G$-instantons. The fundamental results in this regard are two--fold. First,
it was shown by Donaldson and Uhlenbeck and Yau that there is
a one-to-one correspondence between any $G$-instanton solution of the
Hermitian Yang--Mills equation and the existence of a stable holomorphic vector
bundle with structure group $G$ over the Calabi--Yau threefold. Given one the
other is determined, at least in principle. Now, even though it appears to be
very difficult to find solutions of the Hermitian Yang--Mills equations, it
was demonstrated by Friedman, Morgan and Witten~\cite{FMW1,FMW2} and
Donagi~\cite{AJ}
that one can, rather straightforwardly, construct stable holomorphic vector
bundles over Calabi--Yau threefolds. Using, and extending, the technology
introduced in these papers, large classes of heterotic $M$-theory vacua with
non-standard $G$-bundles have been constructed~\cite{gub,smb}. It was shown in these
papers that heterotic $M$-theory vacua corresponding to grand unified theories,
with gauge groups such as $SU(5)$ and $SO(10)$~\cite{gub},
and the standard model with
gauge group $SU(3)_{C} \times SU(2)_{L} \times U(1)_{Y}$~\cite{smb} can, indeed, be
constructed in this manner. I will not discuss these holomorphic bundle
constructions in these lectures, referring the reader to the TASI lectures by
Daniel Waldram. Here, instead, I will discuss an important implication of
non-standard $G$-bundle vacua, namely, the necessary appearance of
$M$5-branes, wrapped on holomorphic curves, in the bulk space.

Recall from above that anomaly cancellation requires that the Bianchi identity
for the four-form field strength $G=dC$ be modified as in
equation~\eqref{Bianchi}. It is useful to rewrite this expression as
\begin{equation}
 (dG)_{11\bar{I}\bar{J}\bar{K}\bar{L}} = -4\sqrt{2}\pi
    \left(\frac{\k}{4\pi}\right)^{2/3} \left\{
       J^{(1)}\d (x^{11}) + J^{(2)}\d (x^{11}-\pi\r )
       \right\}_{\bar{I}\bar{J}\bar{K}\bar{L}}
\label{new13}
\end{equation}
where sources are defined by
\begin{equation}
J^{(n)}= c_{2}(V^{n})-\frac{1}{2}c_{2}(TX) \;\;\;\;\;\; n=1,2,
\label{new23}
\end{equation}
and
\begin{equation}
c_{2}(V^{n})=-\frac{1}{16\pi^{2}}trF^{n} \wedge F^{n}, \qquad
c_{2}(TX)=-\frac{1}{16\pi^{2}}trR \wedge R,
\label{new33}
\end{equation}
$V^{n}$ is the stable holomorphic vector bundle on the $n$-th plane,
$F^{n}$ is the field strength associated with the gauge theory,
and $R$ is the Ricci tensor of the Calabi-Yau manifold.
Note that $c_{2}(V^{n})$ and $c_{2}(TX)$ are the second Chern class of the
vector bundle on the $n$-th boundary plane and the second Chern class of the
Calabi-Yau tangent bundle respectively.
Integrating~\eqref{new13} over a
five-cycle which spans the orbifold interval and is otherwise an arbitrary
four-cycle in the Calabi-Yau three-fold, we find the topological condition that
\begin{equation}
c_{2}(V^{1})+c_{2}(V^{2})-c_{2}(TX)=0.
\label{new43}
\end{equation}

When $N$ bulk five-branes, located at coordinates $x_{i}$ for $i=1,\dots,N$
in the 11-direction, are present in the vacuum, cancellation of their
worldvolume anomalies, as well as the gravitational and gauge anomalies
on the orbifold fixed planes, requires that Bianchi identity be
further modified to
\begin{equation}
(dG)_{11\bar{I}\bar{J}\bar{K}\bar{L}}=4\sqrt{2}\pi\left(\frac{\kappa}
{4\pi}\right)^{2/3}(J^{(1)}\delta(x^{11}) + J^{(2)}\delta(x^{11}-\pi\rho)
+\Sigma_{i=1}^{N} \hat{J}^{(i)}\delta(x^{11}-x_{i}))_{\bar{I}
\bar{J}\bar{K}\bar{L}}.
\label{new53}
\end{equation}
Each five-brane source $\hat{J}^{(i)}$ is defined to be the four-form which is
Poincar\'e dual to the holomorphic curve in the Calabi-Yau threefold
around which the $i$-th five-brane is wrapped. If we define the five-brane class
\begin{equation}
W=\Sigma_{i=1}^{N} \hat{J}^{(i)},
\label{eq:63}
\end{equation}
then the topological condition~\eqref{new43} is modified to
\begin{equation}
c_{2}(V^{1})+c_{2}(V^{2})-c_{2}(TX)+W=0.
\label{new73}
\end{equation}
The simplest example one can present is the
standard embedding, where one fixes the Calabi-Yau three-fold and chooses
the two holomorphic vector bundles so that $V^{1}=TX$ and $V^{2}=0$. It
follows that
\begin{equation}
c_{2}(V^{1})=c_{2}(TX), \qquad c_{2}(V^{2})=0.
\label{new83}
\end{equation}
Note that these Chern classes satisfy the topological condition given
in~\eqref{new73} with
\begin{equation}
W=0.
\label{new94}
\end{equation}
That is, for the standard embedding there are no $M$5--branes in the bulk
space, as we already know from the previous lectures. However, as was shown
in~\cite{gub}, most non-standard $G$-bundles correspond to Chern classes that
require a non-vanishing five-brane class $W$ in order to be anomaly free. In
particular, phenomenologically relevant heterotic $M$-theory vacua, such as
those leading to the standard model gauge group with three families of quarks
and leptons~\cite{smb}, must have bulk five--branes. We will, therefore, spend
the remainder of this lecture discussing the structure and physical properties
of bulk space $M$5--branes wrapped on holomorphic curves.

The inclusion of five-branes in the bulk space not only generalizes the types of
background one can consider, but also introduces new degrees of
freedom into the theory, namely, the dynamical fields on the
five-branes themselves. We will now consider what
low-energy fields survive on one of the five-branes when it is wrapped
around a two-cycle in the Calabi--Yau threefold.

In general, the fields on a single five-brane are as
follows~\cite{GT,KM}. The
simplest are the bosonic coordinates $X^I$ describing the embedding of
the brane into 11-dimensional spacetime. The additional bosonic
field is a world-volume two-form potential $B$ with field strength $H=dB$
satisfying a generalized self-duality condition. For small
fluctuations, the duality condition simplifies to the conventional
constraint $H=*H$. These degrees of freedom are paired with spacetime
fermions $\theta$, leading to a Green--Schwarz type action, with
manifest spacetime supersymmetry and local
kappa-symmetry~\cite{BLNPST,APPS}. (As usual, including the self-dual
field in the action is difficult, but is made possible by either including
an auxiliary field or abandoning a covariant formulation.) For a
five-brane in flat space, one can choose a gauge such that the
dynamical fields fall into a six-dimensional massless tensor multiplet
with $(0,2)$ supersymmetry on the brane
world-volume~\cite{Kallosh,CKvP}. This multiplet has five scalars
describing the motion in directions transverse to the five-brane,
together with the self-dual tensor $H$.

For a five-brane embedded in $S^1/Z_2\times X \times M_4$, to preserve
Lorentz invariance in $M_4$, $3+1$ dimensions of the five-brane must
be left uncompactified. The remaining two spatial dimensions are then
wrapped on a two-cycle of the Calabi--Yau three-fold. To preserve
supersymmetry, the two-cycle must be a holomorphic
curve~\cite{w,bbs,vb}. Thus, from the point of view of a
five-dimensional effective theory on $S^1/Z_2\times M_4$, since two of
the five-brane directions are compactified, it appears as a flat
three-brane (or equivalently a domain wall) located at some point
$x^{11}=x$ on the orbifold. Thus, at low energy, the degrees of
freedom on the brane must fall into four-dimensional supersymmetric
multiplets.

An important question is how much supersymmetry is preserved in the
low-energy theory. One way to address this problem is directly from
the symmetries of the Green--Schwarz action, following the discussion
for similar brane configurations in~\cite{bbs}. Locally, the
11-dimensional spacetime $S^1/Z_2\times X\times M_4$ admits
eight independent Killing spinors $\eta$, so should be described by a
theory with eight supercharges. (Globally, only half of the spinors
survive the non-local orbifold quotienting condition
$\G_{11}\eta(-x^{11})=\eta(x^{11})$, so that, for instance, the
eleven-dimensional bulk fields lead to $N=1$, not $N=2$,
supergravity in four dimensions.) The Green--Schwarz form of the
five-brane action is then invariant under supertranslations generated
by $\eta$, as well as local kappa-transformations. In general the
fermion fields $\theta$ transform as (see for instance
ref.~\cite{CKvP})
\begin{equation}
   \d\theta = \eta + P_+\k
\end{equation}
where $P_+$ is a projection operator. If the brane configuration is
purely bosonic then $\theta=0$ and the variation of the bosonic fields
is identically zero. Furthermore, if $H=0$ then the projection
operator takes the simple form
\begin{equation}
   P_\pm = \frac{1}{2}\left( 1 \pm \frac{1}{6!\sqrt{g}}\e^{m_1\ldots m_6}
          \pt_{m_1}X^{I_1}\dots\pt_{m_6}X^{I_6}\G_{I_1\ldots I_6} \right)
\end{equation}
where $\s^m$, $m=0,\dots ,5$ label the coordinates on the five-brane
and $g$ is the determinant of the induced metric
\begin{equation}
   g_{mn} = \pt_m X^I \pt_n X^J g_{IJ}\; .
\end{equation}

If the brane configuration is invariant for some combination of
supertranslation $\eta$ and kappa-transformation, then we say it is
supersymmetric. Now $\k$ is a local parameter which can be chosen at
will. Since the projection operators satisfy $P_++P_-=1$, we see that
for a solution of $\d\theta=0$, one is required to set $\kappa=-\eta$,
together with imposing the condition
\begin{equation}
   P_-\eta = 0
\label{branesusy}
\end{equation}
For a brane wrapped on a two-cycle in the Calabi--Yau space, spanning
$M_4$ and located at $x^{11}=x$ in the orbifold interval, we
can choose the parameterization
\begin{equation}
   X^\m = \s^{\m}, \qquad
   X^A = X^A(\s,\bar{\s}), \qquad
   X^{11} = x
\end{equation}
where $\s=\s^4+i\s^5$. The condition~\eqref{branesusy} then reads
\begin{equation}
   - \left(i/\sqrt{g}\right) \pt X^A \bar{\pt} X^B \G^{(4)}\G_{AB} \,\eta
        = \eta
\end{equation}
where we have introduced the four-dimensional chirality operator
$\G^{(4)}=\G_0\G_1\G_2\G_3$. Recalling that on the Calabi--Yau three-fold the
Killing spinor satisfies $\G^{\bar{b}}\eta=0$, it is easy to show that
this condition can only be satisfied if the embedding is holomorphic,
that is $X^a=X^a(\s)$, independent of $\bar{\s}$. The condition then
further reduces to
\begin{equation}
   \G^{(4)}\eta = i \eta
\label{4dchiral}
\end{equation}
which, given that the spinor has definite chirality in eleven dimensions as
well as on the Calabi--Yau space, implies that $\G^{11}\eta=\eta$,
compatible with the global orbifold quotient condition. Thus, finally,
we see that only half of the eight Killing spinors, namely those
satisfying~\eqref{4dchiral}, lead to preserved supersymmetries on the
five-brane. Consequently the low-energy four-dimensional theory
describing the five-brane dynamics will have $N=1$ supersymmetry.

The simplest excitations on the five-brane surviving in the low-energy
four-dimensional effective theory are the moduli describing the
position of the five-brane in eleven dimensions. There is a single
modulus $X^{11}$ giving the position of the brane in the orbifold
interval. In addition, there is the moduli space of holomorphic curves
${\cal{C}}_2$ in $X$ describing the position of the brane in the
Calabi--Yau space. This moduli space is generally complicated, and we will
not address its detailed structure here. (As an example,
the moduli space of genus one curves in K3 is K3
itself~\cite{vb}.) However, we note that these moduli are scalars in
four dimensions, and we expect them to arrange themselves as a set of chiral
multiplets, with a complex structure presumably inherited from that of
the Calabi--Yau manifold.

Now let us consider the reduction of the self-dual three-form degrees
of freedom. (Here we are essentially repeating a discussion given
in~\cite{wbranes,KLMVW}.) The holomorphic curve is a Riemann surface
and, so, is characterized by its genus $g$. One recalls that the number
of independent harmonic one-forms on a Riemann surface is given by
$2g$. In addition, there is the harmonic volume two-form
$\Omega$. Thus, if we decompose the five-brane world-volume as
${\cal{C}}_2\times M_4$, we can expand $H$ in zero modes as
\begin{equation}
   H=da\wedge\O+F^u\wedge\l_u+h
\end{equation}
where $\l_u$ are a basis $u=1,\dots ,2g$ of harmonic one-forms on
${\cal{C}}_2$, while the four-dimensional fields are a scalar $a$, $2g$
$U(1)$ vector fields $F^u=dA^u$ and a three-form field strength
$h=db$. However, not all these fields are independent
because of the self-duality condition $H=*H$. Rather, one easily concludes
that
\begin{equation}
   h=*da
\end{equation}
and, hence, that the four-dimensional scalar $a$ and two-form
$b$ describe the same degree of freedom. To analyze the vector
fields, we introduce the matrix ${T_u}^v$ defined by
\begin{equation}
   *\l_u = {T_u}^v\l_v
\end{equation}
If we choose the basis $\l_u$ such that the moduli space metric
$\int_{{\cal{C}}_2}\l_u\wedge (*\l_v)$ is the unit matrix, $T$ is antisymmetric and,
of course, $T^2=-1$. The self-duality constraint implies for the
vector fields that
\begin{equation}
 F^u={T_v}^u*F^v\; .
\end{equation}
If we choose a basis for $F^u$ such that
\begin{equation}
 T={\rm diag}\left(\left(\begin{array}{cc}0&1\\-1&0\end{array}\right),\dots ,
   \left(\begin{array}{cc}0&1\\-1&0\end{array}\right)\right)
\end{equation}
with $g$ two by two blocks on the diagonal, one easily concludes that only
$g$ of the $2g$ vector fields are independent. In conclusion, for a genus
$g$ curve ${\cal{C}}_2$, we have found one scalar and $g$ $U(1)$ vector
fields from the two-form on the five-brane worldvolume. The
scalar has to pair with another scalar to form a chiral $N =1$
multiplet. The only other universal scalar available is the zero mode
of the transverse coordinate $X^{11}$ in the orbifold direction.

Thus, in general, the $N=1$ low-energy theory of a single five-brane
wrapped on a genus $g$ holomorphic curve ${\cal{C}}_2$ has gauge group $U(1)^g$
with $g$ $U(1)$ vector multiplets and a universal chiral multiplet with bosonic
fields $(a,X^{11})$. Furthermore, there is some number of
additional chiral multiplets
describing the moduli space of the curve ${\cal{C}}_2$ in the
Calabi--Yau three-fold.

It is well known that when two regions of the five-brane world-volume in
M--theory come into close proximity, new massless states
appear~\cite{wfq,strom}. These are associated with membranes
stretching between the two nearly overlapping five-brane surfaces.
In general, this can lead to
enhancement of the gauge symmetry. Let us now
consider this possibility, heretofore ignored in our discussion. In
general, one can consider two types of brane degeneracy where parts of
the five-brane world-volumes are in close proximity. The first, and
simplest, is to have $N$ distinct but coincident five-branes, all
wrapping the same cycle ${\cal{C}}_2$ in the Calabi--Yau space and
all located at the same
point in the orbifold interval. Here, the new massless states come from
membranes stretching between the distinct five-brane world-volumes. The
second, and more complicated, situation is where there is a degeneracy
of the embedding of a single five-brane, such that parts of the curve
${\cal{C}}_2$ become close together in the Calabi--Yau space.
In this case, the new
states come from membranes stretching between different parts of the
same five-brane world-volume~\cite{grassi1,grassi2}.
Let us consider these two possibilities separately.

The first case of distinct five-branes is analogous to the M--theory
description of $N$ overlapping type IIB D3-branes, which arise as
$N$ coincident five-branes wrapping the same cycle in a flat torus. In
that case, the $U(1)$ gauge theory on each D3-brane is enhanced to a
$U(N)$ theory describing the full collection of branes. Thus, by
analogy, in our case we would expect a similar enhancement of
each of the $g$ $U(1)$ fields on each five-brane. That is,
when wrapped on a holomorphic curve
of genus $g$, the full gauge group for the low-energy theory
describing $N$ coincident five-branes becomes $U(N)^g$.

The second case is inherently more complicated. It can, however, be clearly
elucidated and studied for Calabi-Yau threefolds which are elliptically fibered.
These manifolds consist of a base two-fold, over any point of which
is fibered an elliptic curve. At almost all points in the base, the
elliptic curve is smooth. However, there is a locus of points, called the
discriminant locus, over which the fibers degenerate. These
degeneracies have specific characteristics and have been classified by
Kodaira~\cite{kod}. If the five-brane is wrapped over a smooth fiber, away
from the discriminant locus, then there are no new massless states. However,
as the fiber approaches the discriminant it degenerates to a specific Kodaira
singularity. Accordingly, the five-brane wrapped on such a fiber begins to
``approach itself'' near the singularity, leading to new, massless states
appearing in the theory. The general theory for computing these massless
states was presented for fibers over both the smooth and singular parts of
discriminant curves in~\cite{grassi1} and~\cite{grassi2} respectively. For
example, consider an elliptically fibered Calabi-Yau threefold over an ${\bf F
\rm}_{3}$ Hirzebruch base and let the five-brane be wrapped on a fiber near a
smooth part of the discriminant curve with Kodaira type $I_{2}$. Then, it was
shown in~\cite{grassi1} that, in addition to the usual states, the
$I_{2}$ degeneracy of the elliptic fiber produces an $SU(2)$ doublet ${\bf 2
\rm}$ of massless $N=2$ hypermultiplets with unit electric charge. In general,
one gets a complicated spectrum of new hypermultiplets and, for sufficiently
intricate Kodaira singularities, new non-Abelian vector multiplets as well.

Summarizing the two cases, we see that for $N$ five-branes wrapping the
same curve ${\cal{C}}_2$ of genus $g$, we expect that the symmetry is enhanced from
$N$ copies of $U(1)^g$ to $U(N)^g$. Alternatively, in the second case,
even for a single brane we can get new massless states if the holomorphic
curve degenerates. These states form hypermultiplets and entended non-Abelian
gauge vector multiplets depending on the exact form of the curve degeneracy.


\section{Lecture 4: Beyond Ho\v rava-Witten Theory }


It is of interest to ask whether one can construct other
orbifolds of $M$-theory beyond the $S^{1}/{\bf Z \rm}_{2}$ example of
\cite{hw1,hw2}. A
first step in this direction was taken by Dasgupta and Muhki \cite{dasmuk}
and Witten \cite{wittens5}
who discussed both local and global anomaly cancellation within the context of
$T^{4}/{\bf Z \rm}_{2}$ orbifolds. A major generalization of these results was
presented in \cite{mlo,phase,new, new2} and \cite{bds,ksty} where all the
$M$-theory orbifolds associated with the
spacetime ${\bf R \rm}^{6} \times K3$ were constructed.
In this fourth lecture, we will, for specificity, consider $M$-theory orbifolds on
$S^{1}/{\bf Z \rm}_{2} \times T^{4}/{\bf Z \rm}_{2}$. It will be demonstrated,
in detail, how such orbifolds can be made anaomaly free, completely
determining both the twisted and untwisted sector spectra in the process,
even on odd dimensional orbifold planes where all anomalies vanish.

The spacetime has topology
${\bf R \rm}^{6} \times S^{1} \times T^{4}$, where each of the
five compact coordinates takes
values on the interval $[-\pi, \pi]$ with the endpoints identified. Let
$x^{\mu}$ parameterize the six non-compact dimensions, while $x^{i}$ and
$x^{11}$ parameterize the $T^{4}$ and $S^{1}$ factors respectively. Then the
${\bf Z \rm}_{2}$ action on $S^{1}$ is defined by
\begin{equation}
\alpha : (x^{\mu}, x^{i}, x^{11})
     \longrightarrow (x^{\mu}, x^{i}, -x^{11})
\label{eq:1n}
\end{equation}
whereas the ${\bf Z \rm}_{2}$ action on $T^{4}$ is
\begin{equation}
\beta : (x^{\mu}, x^{i}, x^{11})
     \longrightarrow (x^{\mu}, -x^{i}, x^{11}) \,.
\label{eq:2n}
\end{equation}
The element $\alpha$ leaves invariant the two ten-planes defined by $x^{11}=0$
and $x^{11}=\pi$, while $\beta$ leaves invariant the sixteen seven-planes
defined when the four coordinates $x^{i}$ individually assume the values $0$
or $\pi$. Finally, $\alpha\beta$ leaves invariant the thirty-two six-planes
defined when all five compact coordinates individually assume the values $0$
or $\pi$. The $\alpha\beta$ six-planes coincide with the intersections of the
$\alpha$ ten-planes with the $\beta$ seven-planes.

A gravitational anomaly arises on each ten-plane due to the coupling of chiral
projections of the bulk gravitino to currents localized on the fixed
planes. Since the two ten-planes are indistinguishable aside from their
position, this anomaly is identical on each of the two planes and can be
computed by conventional means if proper care is used. The reason why extra
care is needed is that each ten-plane anomaly arises from the coupling of
eleven-dimensional fermions to ten-dimensional currents, whereas
standard index theorem results only apply to ten-dimensional fermions
coupled to ten-dimensional currents. If one notes that the index theorem
can be applied to the small radius limit where the two ten-planes
coincide, then the gravitational anomaly on each individual ten-plane can be
computed; it is simply one-half of the index theorem anomaly derived
using the ``untwisted'' sector spectrum in ten-dimensions. By untwisted sector,
we mean the ${\bf Z \rm}_{2}$ projection of the eleven-dimensional bulk space
supergravity multiplet onto each ten-dimension fixed plane. This untwisted spectrum
forms the ten-dimensional $N=1$ supergravity multiplet containing a graviton,
a chiral gravitino, a two-form and a scalar dilaton. We denote by
$R$ the ten-dimensional Riemann tensor, regarded as an $SO(9,1)$-valued form.

As pointed out in \cite{hw1,hw2}, in addition to the
untwisted spectrum, one must allow for the possibility of ``twisted'' sector
$N=1$ supermultiplets that live on each ten-dimensional orbifold plane only. For
the case at hand, the twisted sector spectrum must fall into $N=1$ Yang-Mills
supermultiplets consisting of gauge fields and chiral gauginos.
These will give rise to an additional contribution to the
gravitational anomaly on each ten-plane, as well as to mixed and
pure-gauge anomalies. However, since the twisted sector fields are
ten-dimensional, these anomalies can be computed directly from the standard formulas,
without multiplying by one-half. The twisted sector
gauge group, the dimension of the gauge group and the gauge field strength
on the $i$-th ten-plane are denoted by ${\cal{G}}_{i}$,
$n_{i}={\rm dim}\,{\cal{G}}_{i}$
and $F_{i}$ respectively, for $i=1,2$.

The quantum mechanical one-loop local chiral anomaly on the
$i$-th ten-plane is characterized by the twelve-form
\begin{eqnarray}
I_{12}({\rm 1\!-\!loop})_{i}
     &=& \frac{1}{4}\left(I_{GRAV}^{(3/2)}(R)
     -I_{GRAV}^{(1/2)}(R)\right)
     \nonumber\\[.1in]
     & & \!\! +\frac{1}{2}\left(n_{i}\,I_{GRAV}^{(1/2)}(R)
     +I_{MIXED}^{(1/2)}(R,F_{i})+I_{GAUGE}^{(1/2)}(F_{i})\right)
\label{eq:3n}
\end{eqnarray}
from which the anomaly arises by descent. The constituent polynomials contributing
to the pure gravitational anomaly due to the chiral spin $3/2$ and chiral spin
$1/2$ fermions are
\begin{equation}
I_{GRAV}^{(3/2)}(R)
     =\frac{1}{(2\pi)^{5}6!}\bpl\,
     \frac{55}{56}\,{\rm tr}\,R^{6}
     -\frac{75}{128}\,{\rm tr}\,R^{4}\wedge {\rm tr}\,R^{2}
     +\frac{35}{512}\,({\rm tr}\,R^{2})^{3}\,\bpr
\label{eq:4n}
\end{equation}
and
\begin{equation}
I_{GRAV}^{(1/2)}(R)
     =\frac{1}{(2\pi)^{5}6!}\,\bpl
     -\frac{1}{504}\,{\rm tr}\,R^{6}
     -\frac{1}{384}\,{\rm tr}\,R^{4}\wedge {\rm tr}\,R^{2}
     -\frac{5}{4608}\,(\,{\rm tr}\,R^{2})^{3}\,\bpr
\label{eq:5n}
\end{equation}
respectively, where ${\rm tr}$ is the trace of the $SO(9,1)$ indices.
The polynomials contributing to the mixed and pure-gauge anomalies are due to
chiral spin $1/2$ fermions only and are given by
\brr I_{MIXED}^{(1/2)}(R,F_{i})
     &=& \frac{1}{(2\pi)^{5}6!}\,\bpl\,
     \frac{1}{16}\,{\rm tr}\,R^{4}\wedge {\rm Tr}\,F_{i}^{2}
     +\frac{5}{64}\,(\,{\rm tr}\,R^{2})^{2}\wedge {\rm Tr}\,F_{i}^{2}
     \nonumber\\[.1in]
     & & \hspace{.6in}
     -\frac{5}{8}\,{\rm tr}\,R^{2}\wedge\,{\rm Tr}\,F_{i}^{4}\,\bpr
\label{eq:6n}\err
and
\brr I_{GAUGE}^{(1/2)}(F_{i})
     =\frac{1}{(2\pi)^{5}6!}\,{\rm Tr}\,F_{i}^{6} \,.
\label{eq:7n}\err
Here ${\rm Tr}$ is the trace over the adjoint representation of ${\cal{G}}_{i}$. All
the anomaly polynomials are computed using standard index theorems. Each term
in (\ref{eq:3n}) has a factor of $1/2$ because the relevant fermions are
Majorana-Weyl with half the degrees of freedom of Weyl fermions. The
first two terms in (\ref{eq:3n}) arise from untwisted sector fermions, whereas the
last three terms are contributed by the twisted sector. It follows from the
above discussion that the first two terms must have an
additional factor of 1/2, accounting for the overall coefficient
of $1/4$, whereas the remaining three terms are given exactly by the index theorems.

The quantum anomaly (\ref{eq:3n}) would spoil the consistency of the
theory were it not to cancel against some sort of classical inflow anomaly.
Hence, it is imperative to discern the presence of appropriate
local classical counterterms to cancel against (\ref{eq:3n}).
One begins the analysis of anomaly cancellation by considering
the pure ${\rm tr}\,R^{6}$ term in (\ref{eq:3n}) which is irreducible
and must therefore
identically vanish. It follows from the above that this term is
\brr -\frac{1}{2(2\pi)^{5}6!}\,
     \frac{(n_{i}-248)}{494}\,{\rm tr}\,R^{6} \,.
\label{eq:8n}\err
Therefore, the ${\rm tr}\,R^{6}$ term will vanish if and only if each gauge group
${\cal{G}}_{i}$ satisfies the constraint
\brr n_{i}=248 \,.
\label{eq:9n}\err
Without yet specifying which $248$-dimensional gauge group is permitted, we
substitute $248$ for $n_{i}$ in (\ref{eq:3n}) obtaining
\brr I_{12}({\rm 1\!-\!loop})_{i}
     &=& \frac{1}{2(2\pi)^{5}6!}\,\bpl\,
     -\frac{15}{16}\,{\rm tr}\,R^{4}\wedge {\rm tr}\,R^{2}
     -\frac{15}{64}\,(\,{\rm tr}\,R^{2})^{3}
     +\frac{1}{16}\,{\rm tr}\,R^{4}\wedge {\rm Tr}\,F_{i}^{2}
     \nonumber\\[.1in]
     & & \hspace{.7in}
     +\frac{5}{64}\,(\,{\rm tr}\,R^{2})^{2}\wedge {\rm Tr}\,F_{i}^{2}
     -\frac{5}{8}\,{\rm tr}\,R^{2}\wedge {\rm Tr}\,F_{i}^{4}
     +{\rm Tr}\,F_{i}^{6}\,\bpr \,.
\label{eq:10n}\err
Although non-vanishing, this part of the anomaly is reducible. It follows that
it can be made to cancel as long as it can be factorized into the product
of two terms, a four-form and an eight-form. A necessary requirement for this
to be the case is that
\brr {\rm Tr}\,F_{i}^{6}=
     \frac{1}{24}\,{\rm Tr}\,F_{i}^{4}\wedge {\rm Tr}\,F_{i}^{2}
     -\frac{1}{3600}\,(\,{\rm Tr}\,F_{i}^{2}\,)^{3} \,.
\label{eq:11n}\err
There are two Lie groups with dimension $248$
that satisfy this condition, the non-Abelian group
$E_{8}$ and the Abelian group $U(1)^{248}$. Both groups represent allowed
twisted matter gauge groups on each ten-plane. Hence, from anomaly
considerations alone one can determine the twisted sector on each ten-plane,
albeit with a small ambiguity in the allowed
twisted sector gauge group. In this paper, we consider only the non-Abelian
gauge group $E_{8}$. Using (\ref{eq:11n}) and several $E_{8}$ trace relations,
the anomaly polynomial (\ref{eq:10n}) can be re-expressed as follows
\brr I_{12}({\rm 1\!-\!loop})_{i}=
     \frac{1}{3}\,\pi\,I_{4(i)}^{3}
     + X_{8} \wedge I_{4(i)}
\label{eq:12n}\err
where $X_{8}$ is the eight-form
\brr X_{8}=
     \frac{1}{(2\pi)^{3}4!}\,\bpl\,
     \frac{1}{8}\,{\rm tr}\,R^{4}
     -\frac{1}{32}\,(\,{\rm tr}\,R^{2}\,)^{2}\,\bpr
\label{eq:13n}\err
and $I_{4\,(i)}$ is the four-form given by
\brr I_{4\,(i)}=
    \frac{1}{16\pi^{2}}\,\bpl\,
    \frac{1}{30}\,{\rm Tr}\,F_{i}^{2}
    -\frac{1}{2}\,{\rm tr}\,R^{2}\bpr \,.
\label{eq:14n}\err
Once in this factorized form, the anomaly $I_{12}({\rm 1\!-\!loop})_{i}$
can be cancelled as follows.

First, the Bianchi identity $dG=0$, where G is the field strength of the
three-form $C$ in the eleven-dimensional supergravity multiplet, is modified to
\brr dG=\sum_{i=1}^{2}I_{4(i)} \wedge \delta_{M_{i}^{10}}^{(1)}
\label{eq:15n}\err
where $I_{4(i)}$ is the four-form given in (\ref{eq:14n}) and
$\delta_{M_{i}^{10}}^{(1)}$ is a one-form brane current with support on the
$i$-th ten-plane. Second, we note that the eleven-dimensional supergravity
action contains the terms
\brr S=\cdots -\frac{\pi}{3}
     \int C \wedge G \wedge G+ \int G \wedge X_{7}
\label{eq:16n}\err
where $X_{7}$ satisfies $dX_{7}=X_{8}$.
The $CGG$ interaction is required by the minimally-coupled supergravity
action, while the $GX_{7}$ term  is an additional higher-derivative
interaction necessitated by five-brane anomaly cancellation. Using the
modified Bianchi identity (\ref{eq:15n}), one can compute the variation of
these two terms under Lorentz and gauge transformations. The result is
that the $CGG$ and $GX_{7}$ terms have classical anomalies which descend from
the polynomials
\brr I_{12}(CGG)_{i}=-\frac{\pi}{3}I_{4\,(i)}^{\,3}
\label{eq:17n}\err
and
\brr I_{12}(GX_{7})_{i}=- X_{8} \wedge I_{4(i)} \,.
\label{eq:18n}\err
respectively. It follows that
\brr I_{12}({\rm 1\!-\!loop})_{i}
     +I_{12}(CGG)_{i}
     +I_{12}(GX_{7})_{i}=0
\label{eq:19n}\err
and, hence, the total anomaly cancels exactly.

We conclude that the requirement of local anomaly
cancellation on the each of the two $S^{1}/{\bf Z \rm}_{2}$ orbifold
ten-planes specifies the twisted spectrum of the theory. This specification is
almost, but not quite, unique, allowing $N=1$ vector supermultiplets
with either gauge group $E_{8}$ or $U(1)^{248}$. An important ingredient in
this analysis was the fact that the contribution to the anomaly on each
ten-plane from the untwisted sector was a factor of $1/2$  smaller than the
index theorem result. This followed from the fact that the index theorem
had to be spread over two equivalent ten-planes. A direct consequence of
this is that the non-Abelian gauge group on each ten-plane is $E_{8}$, not
$E_{8} \times E_{8}$, and that the gauge group $SO(32)$
is disallowed. Since
$S^{1}/{\bf Z \rm}_{2}$ is a subspace of $S^{1}/{\bf Z \rm}_{2}
\times T^{4}/{\bf Z \rm}_{2}$, the results of this section continue to hold on
the larger orbifold. We now discuss the cancellation of local anomalies in the
other factor space, $T^{4}/{\bf Z \rm}_{2}$.

The quantum anomalies on each of the sixteen indistinguishable
seven-planes of the $T^{4}/{\bf Z \rm}_{2}$ orbifold are easy to
analyze. In analogy with the ten-planes, an untwisted sector is induced on
each seven-plane by the ${\bf Z \rm}_{2}$ projection of the
eleven-dimensional supergravity  multiplet. This untwisted spectrum forms the
seven-dimensional $N=1$ supergravity multiplet consisting of a graviton,
a gravitino, three vector fields, a two-form, a real scalar dilaton
and a spin 1/2 dilitino.
However, unlike the case of a ten-plane, gravitational anomalies
cannot be supported on a seven-plane. In fact, since there are no chiral fermions in
seven-dimensions, no chiral anomaly of any kind, gravitational or gauge, can arise.
Hence, with no local chiral anomalies to cancel, it would appear to be
impossible to compute the twisted sector spectrum of any seven-plane. As long
as we focus on the seven-planes exclusively, this conclusion is correct. However, as
we will see below, the cancellation of the local anomalies on
the thirty-two six-dimensional $\alpha\beta$ orbifold planes, formed from the
intersection of the $\alpha$ ten-planes with the $\beta$ seven-planes, will
require a non-vanishing twisted sector spectrum on each
seven-plane and dictate its structure. With this in mind, we now turn to
the analysis of anomalies localized on the intersection six-planes in the
full $S^{1}/{\bf Z \rm}_{2} \times T^{4}/{\bf Z \rm}_{2}$ orbifold.

As in the case for the ten-planes, a gravitational anomaly will arise on each
six-plane due to the coupling of chiral projections of the bulk gravitino to
currents localized on the thirty-two fixed planes. Since the thirty-two
six-planes are indistinguishable, the anomaly is the same on each
plane and can be computed by conventional means if proper care is
taken. Noting that the standard index theorems can be applied to the small
radius limit where the thirty-two six-planes coincide, it follows that the
gravitational anomaly on each six-plane is simply one-thirty-second of
the index theorem anomaly derived using the untwisted sector spectrum in
six-dimensions. In this case, the untwisted sector spectrum  is the
${\bf Z \rm}_{2} \times {\bf Z \rm}_{2}$ projection of the
eleven-dimensional bulk supergravity
multiplet onto each six-dimensional fixed plane. This untwisted spectrum forms
several $N=1$ six-dimensional supermultiplets. Namely,
the supergravity multiplet consisting
of a graviton, a chiral gravitino and a self-dual two-form,
four hypermultiplets each with four
scalars and an anti-chiral hyperino,
and one tensor multiplet with one anti-self-dual
two-form, one scalar and an anti-chiral spin 1/2 fermion.
A one-loop quantum
gravitational anomaly then arises from one chiral spin $3/2$ fermion, five
anti-chiral spin $1/2$ fermions and one each of self-dual and anti-self-dual
tensors. However, the anomalies due to the tensors cancel each other. Noting
that a chiral anomaly in six-dimensions is characterized by an eight-form,
from which the anomaly arises by descent, we find, for the $i$-th six-plane,
that
\brr I_{8}(SG)_{i}=
     \frac{1}{32}\,\bpl\,
     I_{GRAV}^{(3/2)}(R)
     -5\,I_{GRAV}^{(1/2)}(R)\,\bpr
\label{eq:20n}\err
where
\brr I_{GRAV}^{(3/2)}(R)=
     \frac{1}{(2\pi)^{3}4!}\,\bpl
     -\frac{49}{48}\,{\rm tr}\,R^{4}
     +\frac{43}{192}\,(\,{\rm tr}\,R^{2}\,)^{2}\,\bpr
\label{eq:21n}\err
and
\brr I_{GRAV}^{(1/2)}(R)=
    \frac{1}{(2\pi)^{3}4!}\,\bpl
    -\frac{1}{240}\,{\rm tr}\,R^{4}
    -\frac{1}{192}\,(\,{\rm tr}\,R^{2})^{2}\,\bpr \,,
    \label{eq:22n}\err
where $R$ is the six-dimensional Riemann tensor, regarded as an $SO(5,1)$-valued
form. Note that the terms in brackets in (\ref{eq:20n}) are the anomaly as
computed by the index theorem.
$I_{8}(SG)_{i}$ is obtained from that result by dividing by $32$.

Noting that each six-plane is embedded in one of the two ten-dimensional
planes, we see that there are additional ``untwisted'' sector fields on each
six-plane. These arise from the $\beta$
${\bf Z \rm}_{2}$ projection of the $N=1$ $E_{8}$ Yang-Mills supermultiplet on
the associated ten-plane. Such fields are untwisted from the point of view of
the six-dimensional plane, although they arise from fields that were in the
twisted sector of the ten-plane. In this lecture, we will assume
that the $\beta$ action on the ten-dimensional vector multiplets does not break
the $E_{8}$ gauge group. A discussion of the case where $E_{8}$ is broken to
a subgroup by the action of $\beta$ can be found in \cite{phase,new}.
A ten-dimensional
$N=1$ vector supermultiplet decomposes in six-dimensions into an $N=1$ vector
multiplet and an $N=1$ hypermultiplet. However, the action of $\beta$ projects
out the hypermultiplet. Therefore, the ten-plane contribution to the untwisted
sector of each six-plane is an $N=1$ $E_{8}$ vector supermultiplet, which
consists of gauge fields and chiral gauginos. The gauginos contribute to the
gravitational anomaly on each six-plane, as well as adding mixed and
$E_{8}$ gauge anomalies. Noting that the standard index theorems can be
applied to the small radius limit, where each ten-plane shrinks to zero size
and, hence,  the sixteen six-planes it contains coincide,
it follows that the anomaly is
simply one-sixteenth of the index theorem result. We find that the
one-loop quantum contribution of this $E_{8}$ supermultiplet to the
gravitational, mixed and $E_{8}$ gauge anomalies on the $i$-th six-plane is
\brr I_{8}(E_{8})_{i}=
     \frac{1}{16}\,\bpl\,248\,I_{GRAV}^{(1/2)(R)}
     +I_{MIXED}^{(1/2)}(R,F_{i})
     +I_{GAUGE}^{(1/2)}(F_{i})\,\bpr
\label{eq:23n}\err
where
\brr I_{MIXED}^{(1/2)}(R,F_{i})=
     \frac{1}{(2\pi)^{3}4!}\,\bpl\,
     \frac{1}{4}\,{\rm tr}\,R^{2}\wedge {\rm Tr}\,F_{i}^{2}\,\bpr
\label{eq:24n}\err
and
\brr I_{GAUGE}^{(1/2)}(F_{i})=
     \frac{1}{(2\pi)^{3}4!}\,\bpl
     -{\rm Tr}\,F_{i}^{4}\,\bpr \,.
\label{eq:25n}\err
Here $\rm Tr$ is over the adjoint ${\bf 248 \rm}$ representation of $E_{8}$.
Note that the terms in brackets in (\ref{eq:23n}) are the index theorem
anomaly. $I_{8}(E_{8})_{i}$ is obtained from that result by dividing by $16$.

Are there other sources of untwisted sector anomalies on a six-plane? The
answer is, potentially yes. We note that, in addition to being embedded in one
of the two ten-planes, each six-plane is also embedded in one of the sixteen
seven-dimensional orbifold planes. In analogy with the discussion above, if
there were to be a non-vanishing twisted sector spectrum on each seven-plane,
then this could descend under the $\alpha$ ${\bf Z \rm}_{2}$ projection as an
addition to the untwisted spectrum on each six-plane. This additional
untwisted spectrum could then contribute to the chiral anomalies on the
six-plane. However, as  noted above, a priori,
there is no reason for one
to believe that there is any twisted sector on a seven-dimensional orbifold
plane. Therefore, for the time being, let us assume that there is no such
contribution to the six-dimensional anomaly. We will see below that this
assumption must be carefully revisited.

As for the ten-dimensional planes, one must allow for the
possibility of twisted sector $N=1$ supermultiplets on each of the
thirty-two six-planes. The most general allowed spectrum on the $i$-th
six-plane  would be $n_{Vi}$ vector multiplets
transforming in the adjoint
representation of some as yet unspecified gauge group ${\cal{G}}_{i}$, $n_{Hi}$
hypermultiplets transforming under some representation (possibly reducible)
${\cal{R}}$ of ${\cal{G}}_{i}$, and $n_{Ti}$ gauge-singlet tensor multiplets.
We denote by ${\cal{F}}_{i}$ the gauge field strength.
Since these fields are in the twisted sector, their contribution to the chiral
anomalies can be determined directly from the index theorems without
modification. We find that the one-loop quantum contribution of the twisted
spectrum to the gravitational, mixed and ${\cal{G}}_{i}$ gauge anomalies on
the $i$-th six-plane is
\brr I_{8}({\cal{G}}_{i})
     &=& (n_{V}-n_{H}-n_{T})_{i}\,I_{GRAV}^{(1/2)}(R)
     -n_{Ti}\,I_{GRAV}^{({\rm 3\!-\!form})}(R)
     \nonumber\\[.1in]
     & & +I_{MIXED}^{(1/2)}(R,{\cal{F}}_{i})
     +I_{GAUGE}^{(1/2)}({\cal{F}}_{i})
\label{eq:26n}\err
where $I_{GRAV}^{(1/2)}(R)$ is given in (\ref{eq:22n}) and
\brr I_{GRAV}^{({\rm 3\!-\!form})}(R)=
    \frac{1}{(2\pi)^{3}4!}\,\bpl
    -\frac{7}{60}\,{\rm tr}\,R^{4}
    +\frac{1}{24}\,(\,{\rm tr}\,R^{2}\,)^{2}\,\bpr \,.
\label{eq:27n}\err
Furthermore, the mixed and pure-gauge anomaly polynomials are modified to
\brr I_{MIXED}^{(1/2)}(R,{\cal{F}}_{i})=
     \frac{1}{(2\pi)^{3}4!}\,\bpl\,
     \frac{1}{4}\,{\rm tr}\,R^{2}\wedge
     {\rm trace}\,{\cal{F}}_{i}^{2}\,\bpr
\label{eq:28n}\err
and
\brr I_{GAUGE}^{(1/2)}({\cal{F}}_{i})=
     \frac{1}{(2\pi)^{3}4!}\,\bpl
     -{\rm trace}\,{\cal{F}}_{i}^{4}\,\bpr \,,
\label{eq:29n}\err
where
\brr {\rm trace}\,{\cal{F}}_{i}^{n}=
     {\rm Tr}\,{\cal{F}}_{i}^{n}
     -\sum_{\alpha}\,h_{\alpha}\,{\rm tr}_{\alpha}{\cal{F}}_{i}^{n} \,.
\label{eq:30n}\err
Here $\rm Tr$ is an adjoint trace, $h_{\alpha}$ is the number of hypermultiplets
transforming in the ${\cal{R}}_{\alpha}$ representation and $tr_{\alpha}$ is a
trace over the ${\cal{R}}_{\alpha}$ representation. Note that the total number
of vector multiplets is $n_{Vi}={\rm dim}\,({\cal{G}}_{i})$,
while the total number of
hypermultiplets is
$n_{Hi}=\sum_{\alpha}h_{\alpha} \times {\rm dim}\,({\cal{R}}_{\alpha})$.
The relative minus sign in (\ref{eq:30n}) reflects
the anti-chirality of the hyperinos.

Combining the contributions from the two untwisted sector sources and the
twisted sector, the total one-loop quantum anomaly on the $i$-th six-plane is
the sum
\brr I_{8}({\rm 1\!-\!loop})_{i}=
     I_{8}(SG)_{i}+I_{8}(E_{8})_{i}
     +I_{8}({\cal{G}}_{i})
\label{eq:31n}\err
where $I_{8}(SG)_{i}$, $I_{8}(E_{8})_{i}$ and $I_{8}({\cal{G}}_{i})$ are given
in (\ref{eq:20n}), (\ref{eq:23n}) and (\ref{eq:26n}) respectively.

Unlike the case for the ten-dimensional planes, the classical anomaly
associated with the $GX_{7}$ term in the eleven-dimensional action
(\ref{eq:16n}) can contribute to the irreducible curvature term which, in
six-dimensions, is ${\rm tr}\,R^{4}$. Therefore, our
next step is to further modify the Bianchi identity for $G=dC$ from expression
(\ref{eq:15n}) to
\brr dG=\sum_{i=1}^{2}I_{4(i)} \wedge \delta_{M_{i}^{10}}^{(1)}
     +\sum_{i=1}^{32}g_{i}\,\delta_{M^{6}_{i}}^{(5)}
\label{eq:32n}\err
where $\delta_{M^{6}_{i}}^{(5)}$ has support on the six-planes
$M^{6}_{i}$. As discussed in \cite{mlo,phase}, the magnetic charges
$g_{i}$ are required to take the values
\brr g_{i}=-3/4, -1/4, +1/4,...
\label{eq:33n}\err
Using the modified Bianchi identity (\ref{eq:31n}), one can compute the
variation of the $GX_{7}$ term under Lorentz  and gauge
transformations. The result is that this term gives rise to a classical
anomaly that descends from the polynomial
\brr I_{8}(GX_{7})_{i}=-g_{i}\,X_{8}
\label{eq:35n}\err
where $X_{8}$ is presented in expression (\ref{eq:13n}).
The relevant anomaly is then
\brr I_{8}({\rm 1\!-\!loop})_{i}
     +I_{8}(GX_{7})_{i}
\label{eq:36n}\err
where $I_{8}({\rm 1\!-\!loop})_{i}$ is given in (\ref{eq:31n}).
This anomaly spoils the consistency of the theory and, hence, must cancel. One
begins the analysis of anomaly cancellation by considering the pure
${\rm tr}\,R^{4}$
term in (\ref{eq:36n}) which is irreducible and must identically vanish.
It follows from the above that this term is
\brr -\frac{1}{(2\pi)^{3}4!\,240}\,
    (n_{Vi}-n_{Hi}-29n_{Ti}
    +30\,g_{i}+23)\,{\rm tr}\,R^{4} \,.
\label{eq:37n}\err
Therefore, the ${\rm tr}\,R^{4}$ term will vanish if and only if on each orbifold
plane the constraint
\brr n_{Vi}-n_{Hi}-29n_{Ti}+30g_{i}+23=0
\label{eq:38n}\err
is satisfied. Herein lies a problem, and the main point of paper~\cite{new2}.
Noting from (\ref{eq:33n}) that $g_{i}=c_{i}/4$ where $c_{i}=-3,-1,1,3,5,...$,
we see that cancelling the ${\rm tr}\,R^{4}$ term requires that we satisfy
\brr n_{Vi}-n_{Hi}-29n_{Ti}=(-15c_{i}-46)/2 \,.
\label{eq:39n}\err
However, this is not possible since the left hand side of this expression
is an integer and the right hand side always half integer. There is only one
possible resolution of this problem, which is to carefully review the
only assumption that was made above, that is, that
there is no twisted sector on a seven-plane and,
hence, no contribution of the seven-planes by $\alpha$ ${\bf Z \rm}_{2}$
projection to the untwisted anomaly on a six-plane. As we now show,
this assumption is false.

Let us now allow for the possibility that there is a twisted sector of $N=1$
supermultiplets on each of the sixteen seven-planes. The most general allowed
spectrum on the $i$-th seven-plane would be $n_{7Vi}$
vector supermultiplets transforming in the adjoint representation of some as yet
unspecified gauge group $G_{7i}$. Each seven-dimensional vector
multiplet contains a gauge field, three scalars and a gaugino. With respect to
six-dimensions, this vector multiplet decomposes into an $N=1$ vector
supermultiplet and a single hypermultiplet. Under the
$\alpha$ ${\bf Z \rm}_{2}$ projection to each of the two embedded six-planes,
the gauge group $G_{7i}$ can be preserved or broken to a subgroup. In
either case, we denote the six-dimensional gauge group arising in this manner
as $\tilde{\cal{G}}_{i}$, define $\tilde{n}_{Vi}={\rm dim}\,\tilde{\cal{G}}_{i}$ and
write the associated gauge field strength as $\tilde{\cal{F}}_{i}$.
In this lecture, for simplicity, we will assume that the
gauge group is unbroken by the orbifold projection, that is,
$\tilde{\cal{G}}_{i}=G_{7i}$. The more general case where it is broken to a
subgroup is discussed in \cite{phase,new}. Furthermore, the $\alpha$ action
projects out either the six-dimensional vector supermultiplet, in which case
the hypermultiplet descends to the six-dimensional untwisted sector,
or the six-dimensional hypermultiplet, in which case the vector supermultiplet
enters the six-dimensional untwisted sector. We denote by $\tilde{n}_{Hi}$ the
number of hypermultiplets arising in the six-dimensional untwisted sector
by projection from the seven-plane, and specify their (possibly reducible)
representation
under $\tilde{\cal{G}}_{i}$ as $\tilde{\cal{R}}$. Since these fields are in
the untwisted sector associated with a single seven-plane, and since there are
two six-planes embedded in each seven-plane, their contribution to the quantum
anomaly on each six-plane can be determined by taking $1/2$ of the index
theorem result. We find that the one-loop quantum contribution of this part of
the the untwisted spectrum to the gravitational, mixed and $\tilde{\cal{G}}_{i}$
gauge anomalies on the $i$-th six-plane is
\brr I_{8}(\tilde{\cal{G}}_{i})=
     \frac{1}{2}\,\bpl\,
     (\tilde{n}_{V}-\tilde{n}_{H})_{i}\,I_{GRAV}^{(1/2)}(R)
     +I_{MIXED}^{(1/2)}(R,\tilde{{\cal{F}}}_{i})
     +I_{GAUGE}^{(1/2)}(\tilde{{\cal{F}}}_{i})\,\bpr
\label{eq:40n}\err
where $I_{GRAV}^{(1/2)}(R), I_{MIXED}^{(1/2)}(R,\tilde{{\cal{F}}}_{i})$
and $I_{GAUGE}^{(1/2)}(\tilde{{\cal{F}}}_{i})$
are given in (\ref{eq:22n}),(\ref{eq:28n}) and (\ref{eq:29n}) respectively with
the gauge and hypermultiplet quantities replaced by their `` $\sim$ ''
equivalents.

The total quantum anomaly on the $i$-th six-plane is now modified to
\brr I_{8}({\rm 1\!-\!loop})_{i}
     +I_{8}(\tilde{\cal{G}}_{i})
\label{eq:41n}\err
where $I_{8}({\rm 1\!-\!loop})_{i}$ and $I_{8}(\tilde{\cal{G}}_{i})$ are given
in (\ref{eq:31n}) and (\ref{eq:40n}) respectively. It follows that the relevant
anomaly contributing to, among other things, the irreducible ${\rm tr}\,R^{4}$ term
is modified to
\brr I_{8}({\rm 1\!-\!loop})_{i}
     +I_{8}(\tilde{\cal{G}}_{i})
     +I_{8}(GX_{7})_{i} \,.
\label{eq:42n}\err
This anomaly spoils the quantum consistency of the theory and, hence, must
cancel. We again begin by considering the pure ${\rm tr}\,R^{4}$ term in
(\ref{eq:42n}).
This term is irreducible and must identically vanish. It follows from the
above that this term is
\brr -\frac{1}{(2\pi)^{3}4!\,240}\,
    (\,n_{Vi}-n_{Hi}
    +\ft12\,\tilde{n}_{Vi}
    -\ft12\,\tilde{n}_{Hi}
    -29n_{Ti}+30g_{i}+23\,)\,{\rm tr}\,R^{4} \,.
\label{eq:43n}\err
Therefore, the ${\rm tr}\,R^{4}$ term will vanish if and only if on each orbifold
plane the constraint
\brr n_{Vi}-n_{Hi}
    +\ft12\,\tilde{n}_{Vi}-\ft12\,\tilde{n}_{Hi}
    -29\,n_{Ti}+30\,g_{i}+23=0
\label{eq:44n}\err
is satisfied. Again, noting that $g_{i}=c_{i}/4$ where $c_{i}=-3,-1,1,3,5,...$,
we see that we must satisfy
\brr n_{Vi}-n_{Hi}
     +\ft12\,\tilde{n}_{Vi}-\ft12\,\tilde{n}_{Hi}
     -29\,n_{Ti}=\ft12\,(\,-15c_{i}-46\,) \,.
\label{eq:45n}\err
As above, the right hand side is always a half integer. Now, however,
because of the addition of the untwisted spectrum arising from the
seven-plane, the left hand side can also be chosen to be half integer. Hence,
the pure ${\rm tr}\,R^{4}$ term can be cancelled.

Having cancelled the irreducible ${\rm tr}\,R^{4}$ term, we now compute the
remaining terms in the anomaly eight-form. In addition to the contributions
from (\ref{eq:42n}), we must also take into account the classical anomaly
associated with the $CGG$ term in the eleven-dimensional action (\ref{eq:16n}).
Using the modified Bianchi identity (\ref{eq:31n}), one can compute the
variation of the $CGG$ term under Lorentz and gauge transformations. The
result is that this term gives rise to a classical anomaly that descends from
the polynomial
\brr I_{8}(CGG)_{i}=
     -\pi\,g_{i}\,I_{4\,(i)}^{\,2}
\label{eq:46n}\err
where $I_{4\,(i)}$ is given in expression (\ref{eq:14n}). Adding this anomaly to
(\ref{eq:42n}), and cancelling the ${\rm tr}\,R^{4}$ term by imposing constraint
(\ref{eq:44n}), we can now determine the remaining terms in the anomaly
eight-form.

Recall that, in this lecture, we are assuming that the $\beta$ action on the
ten-dimensional vector supermultiplet does not break the $E_{8}$ gauge group.
In this case, we can readily show that there can be no twisted sector vector
multiplets on any six-plane. Rather than complicate the present discussion, we
will simply assume here that gauge field strengths ${\cal{F}}_{i}$ do not
appear. Furthermore, cancellation of the
complete anomaly, in the case where $E_{8}$ is unbroken, requires that
$\tilde{{\cal{G}}}_{i}$ be a product of $U(1)$ factors. Here, we will limit
the discussion to the simplest case where
\brr \tilde{{\cal{G}}}_{i}=U(1)
\label{eq:an}\err
The $\beta$ action on the seven-dimensional plane then
either projects a single vector supermultiplet, or a single
chargeless hypermultiplet,
onto the untwisted sector of the six-plane. In either case, no $U(1)$ anomaly
exists. Hence, the gauge field strengths $\tilde{{\cal{F}}}_{i}$ also do not
appear. With this in mind, we now compute the remaining terms in the anomaly
eight-form. They are
\brr & & \frac{1}{(2\pi)^{3}4!\,16}\,\bpl\,
     \ft34\,(\,1-4\,n_{Ti}\,)\,({\rm tr}\,R^{2})^{2}
     +\ft{1}{20}\,(\,5+8\,g_{i}\,)\,
     {\rm tr}\,R^{2}\wedge{\rm Tr}\,F_{i}^{2}
     \nonumber\\[.1in]
     & & \hspace{.8in}
     -\ft{1}{100}\,(\,1+\ft43\,g_{i}\,)\,
     (\,{\rm Tr}\,F_{i}^{2})^{2}\,\bpr
\label{eq:47n}\err
where we have used the $E_{8}$ trace relation
${\rm Tr}\,F^{4}=\frac{1}{100}({\rm Tr}\,F^{2})^{2}$.
Note that, since $n_{Ti}$ is a non-negative integer and $g_{i}$ must satisfy
(\ref{eq:33n}), the first two terms of this expression term can never vanish.
Furthermore, it is straightforward to show that (\ref{eq:47n}) will
factor into an exact square, and, hence, be potentially cancelled by a
six-plane Green-Schwarz mechanism, if and only if
\brr 4\,(\,4\,n_{Ti}-1\,)(\,3+4\,g_{i}\,)=(\,5+8\,g_{i}\,)^{2}
\label{eq:48n}\err
Again, this equation has no solutions for the allowed values of $n_{Ti}$ and
$g_{i}$. It follows that anomaly (\ref{eq:47n}), as it presently stands, cannot
be be made to identically vanish or cancel. The resolution of this problem
was first described in \cite{mlo}, and
consists of the realization that the existence of seven-planes in
the theory necessitates the introduction of additional Chern-Simons
interactions in the action, one for each seven-plane. The required terms are
\brr S=\cdots
     +\sum_{i=1}^{16}\int
     \delta_{M_{i}^{7}}^{(4)}\wedge G \wedge Y_{3(i)}^{0}
\label{eq:49n}\err
where $dY_{3(i)}^{0}=Y_{4(i)}$ is a gauge-invariant four-form polynomial.
$Y_{4(i)}$ arises from the curvature $R$ and also the field strength
$\tilde{\cal{F}}_{i}$ associated with the additional adjoint
super-gauge fields localized on the $i$-th seven-plane. It is given by
\brr Y_{4(i)}=
     \frac{1}{4\pi}\bpl
     -\frac{1}{32}\,\eta\,{\rm tr}\,R^{2}
     +\rho\,{\rm tr}\,\tilde{\cal{F}}_{i}\,\bpr
\label{eq:50n}\err
where $\eta$ and $\rho$ are rational coefficients. Using the modified Bianchi
identity (\ref{eq:32n}), one can compute the variation of the
$\delta^{7}GY_{3}$ terms under Lorentz and gauge transformations. The
result is that these give rise to a classical anomaly that descends from the
polynomial
\brr I_{8}(\delta^{7}GY_{3})_{i}=
     -I_{4\,(i)} \wedge Y_{4(i)}
\label{eq:51n}\err
where $I_{4\,(i)}$ is the four-form given in (\ref{eq:14n}).

The total anomaly on the $i$-th six-plane is now modified to
\brr I_{8}({\rm 1\!-\!loop})_{i}
     +I_{8}(\tilde{\cal{G}}_{i})
     +I_{8}(GX_{7})_{i}
     +I_{8}(CGG)_{i}
     +I_{8}(\delta^{7}GY_{3})_{i}
\label{eq:52n}\err
where $I_{8}({\rm 1\!-\!loop})_{i}, I_{8}(\tilde{\cal{G}}_{i}),
I_{8}(GX_{7})_{i}, I_{8}(CGG)_{i}$ and $I_{8}(\delta^{7}GY_{3})_{i}$
are given in (\ref{eq:31n}), (\ref{eq:40n}), (\ref{eq:35n}), (\ref{eq:46n}) and
(\ref{eq:51n}) respectively. Note that for the fixed plane intersection
presently under discussion, the field strength $\tilde{\cal{F}}_{i}$ does
not enter the anomaly eight-form (\ref{eq:47n}). Therefore, within this context,
we must take
\brr \rho=0 \,.
\label{eq:addn}\err
After cancelling the irreducible ${\rm tr}\,R^{4}$ term,
the remaining anomaly now becomes
\brr & & \frac{1}{(2\pi)^{3}4!16}\bpl\,
     \frac{3}{4}\,(1-4n_{Ti}-\eta)\,({\rm tr}\,R^{2})^{2}
     \nonumber\\[.1in]
     & & \hspace{.8in}
     +\frac{1}{20}\,(5+8g_{i}+\eta)\,{\rm tr}\,R^{2}\wedge {\rm Tr}\,F_{i}^{2}
     -\frac{1}{100}\,(1+\frac{4}{3}g_{i})\,({\rm Tr}\,F_{i}^{2})^{2}\,\bpr
\label{eq:53n}\err
Depending on the number of untwisted hypermultiplets, $n_{Ti}$, these terms
can be made to cancel or to factor into the sum of exact squares. In
this lecture, we consider the $n_{Ti}=0,1$ cases only. As discussed in
\cite{phase,new}, the solutions where $n_{Ti}\geq2$ are related to the
$n_{Ti}=0,1$ solutions by the absorption of one or more five-branes
from the bulk space onto the $i$-th six-plane.

We first consider the case where
\brr n_{Ti}=0 \,.
\label{eq:bn}\err
In this case, no further Green-Schwarz type mechanism
in six-dimensions is possible and the anomaly must vanish identically. We see
from (\ref{eq:53n}) that this is possible if and only if
\brr g_{i}=-3/4, \qquad \eta=1 \,.
\label{eq:54n}\err
It is important to note that this solution only exists
for a non-vanishing value of parameter $\eta$. Hence, the additional
Chern-Simons interactions (\ref{eq:49n}) are essential for the anomaly
to vanish identically in the $n_{Ti}=0$ case.
Inserting these results into expression (\ref{eq:44n}) for the vanishing of the
irreducible ${\rm tr}\,R^{4}$ term, and recalling that $n_{Vi}=0$, we find that
\brr -2n_{Hi}+\tilde{n}_{Vi}-\tilde{n}_{Hi}=-1 \,.
\label{eq:55n}\err
Equation (\ref{eq:55n}) can be solved in several ways.
Remembering that $\tilde{{\cal{G}}}_{i}=U(1)$,
the first solution then consists of allowing the $U(1)$ hypermultiplet to
descend to the six-plane while projecting out the $U(1)$ vector multiplet.
Equation (\ref{eq:55n}) is then solved by taking the number of twisted
hypermultiplets to vanish. That is, take
\brr \tilde{n}_{Hi}=1, \qquad \tilde{n}_{Vi}=0, \qquad n_{Hi}=0 \,.
\label{eq:56n}\err
The second solution follows by doing the reverse, that is, projecting out the
$U(1)$ hypermultiplet and allowing the $U(1)$ vector multiplet to descend to
the six-plane. In this case, equation (\ref{eq:55n}) is solved by taking
\brr \tilde{n}_{Hi}=0, \qquad \tilde{n}_{Vi}=1, \qquad n_{Hi}=1 \,.
\label{eq:57n}\err

Let us now consider the case where
\brr n_{Ti}=1 \,.
\label{eq:58n}\err
In this case, the anomaly
(\ref{eq:53n}) can be removed by a six-dimensional Green-Schwarz mechanism
as long as it factors into an exact square. It is straightforward to show that
this will be the case if and only if
\brr 4\,(\,3+\eta\,)\,(\,3+4\,g_{i}\,)=(\,5+8\,g_{i}+\eta\,)^{2} \,.
\label{eq:59n}\err
This equation has two solutions
\brr g_{i}=-3/4, \qquad \eta=1
\label{eq:60n}\err
and
\brr g_{i}=1/4, \qquad \eta=1 \,.
\label{eq:61n}\err
Again, note that these solutions require a non-vanishing value of the
parameter $\eta$. Hence, the additional Chern-Simons interactions
(\ref{eq:49n}) are also essential for anomaly factorization in the $n_{Ti}=1$
case.
Inserting these into the expression for the vanishing of the irreducible
${\rm tr}\,R^{4}$ term, and recalling that $n_{Vi}=0$, we find
\brr -2n_{Hi}+\tilde{n}_{Vi}-\tilde{n}_{Hi}=57
\label{eq:62n}\err
and
\brr -2n_{Hi}+\tilde{n}_{Vi}-\tilde{n}_{Hi}=-3 \,.
\label{eq:63n}\err
The first equation (\ref{eq:62n}) cannot be solved within the context of
$\tilde{{\cal{G}}}_{i}=U(1)$, since $\tilde{n}_{Vi} \leq 1$. The second equation,
however, has two solutions
\brr \tilde{n}_{Hi}=1, \qquad \tilde{n}_{Vi}=0, \qquad n_{Hi}=1
\label{eq:64n}\err
and
\brr \tilde{n}_{Hi}=0, \qquad \tilde{n}_{Vi}=1, \qquad n_{Hi}=2 \,.
\label{eq:65n}\err

In either case, the anomaly (\ref{eq:53n}) factors into an exact square given
by
\brr -\frac{3}{(2\pi)^{3}4!\,16}\,\bpl
     {\rm tr}\,R^{2}
     -\frac{1}{15}\,{\rm Tr}\,F_{i}^{2}\,\bpr^2 \,.
\label{eq:66n}\err
The anomaly can now be cancelled by a Green-Schwarz mechanism on the
six-plane. First, one alters the Bianchi identity for the anti-self-dual tensor
in the twisted sector tensor multiplet
from $dH_{Ti}=0$, where $H_{Ti}$ is the tensor field strength three-form, to
\brr dH_{Ti}=\frac{1}{16\pi^{2}}
    (\,{\rm tr}\,R^{2}-\frac{1}{15}{\rm Tr}\,F_{i}^{2}\,) \,.
\label{eq:67n}\err
Second, additional Chern-Simons terms are added to the action, one for each
six-plane. The required terms are
\brr S=\cdots-\frac{1}{64\pi}
    \sum_{i=1}^{32}\int \delta^{(5)}_{M_{i}^{6}}
    \wedge B_{Ti} \wedge (\,{\rm tr}\,R^{2}-\frac{1}{15}{\rm Tr}\,F_{i}^{2}\,) \,,
\label{eq:68n}\err
where $B_{Ti}$ is the anti-self-dual tensor two-form on the $i$-th six-plane.
Using Bianchi identity (\ref{eq:67n}), one can compute the variation of each
such term under Lorentz and gauge transformations. The result is a classical
anomaly that descends from an eight-form that exactly cancels
expression (\ref{eq:66n}). The theory is now anomaly free.

Thus, we have demonstrated, within the context of an explicit orbifold
fixed plane intersection where the $\beta$ ${\bf Z \rm}_{2}$ projection to the
six-plane leaves $E_{8}$ unbroken, that all local anomalies
can be cancelled.
However, this cancellation requires that the intersecting
seven-plane support a
twisted sector consisting of a $U(1)$ $N=1$ vector supermultiplet and an
associated Chern-Simons term. This term is of the form (\ref{eq:49n}) with
$\eta=1$ and $\rho=0$. The fact that $\rho=0$ in this context
follows directly from the
property that $E_{8}$ is unbroken by the $\beta$ projection.

We conclude that, as has been discussed in detail in~\cite{mlo,phase,new, new2}
and~\cite{bds,ksty}, anomaly free $M$-theory orbifolds associated with the
spacetime ${\bf R \rm}^{6} \times K3$ can be constructed in detail, including
the entire twisted and untwisted spectra. This work has now been extended to
orbifolds of spacetime ${\bf R \rm}^{4} \times CY_{3}$, where $CY_{3}$ is a
Calabi-Yau threefold, in~\cite{mike2}. This last work opens the door to finding
realistic standard model-like $M$-theory vacua within this context.


\section{ Discussion }


Ho\v rava-Witten theory and its compactification on Calabi-Yau threefolds to
heterotic $M$-theory have stimulated a great deal of both formal $M$-theory
research as well as discussions of the associated phenomenology. In addition to
the papers referenced in the above lectures, further relevant literature can
be found
in~\cite{aq1,kap,ns,lln,conrad,du1,dm,llo,lpt,lt,du2,efn,nan,ms,nw,bkl,
sharpe,pes,elpp,deA}.

Heterotic $M$-theory has also served as a consistent and phenomenologically
relevant venue for studying $M$-theory cosmology. This research comes in two
catagories. The first consists of work discussing subluminal expansion,
Kasner-like solutions and inflation within the context of brane world scenarios
associated with heterotic $M$-theory. These results can be found in~\cite{lo,
benakli,low4,low6,low7,hos,blo}. Very recently, a new theory of the early
universe, called the Ekpyrotic Universe, has been constructed for generic
brane world scenarios, including heterotic $M$-theory. In the Ekpyrotic
scenario, all expansion is subluminal, with no period of inflation. A nearly
scale-invariant spectrum of fluctuations in the microwave background is
obtained, not as quantum fluctuations in deSitter space but, rather, as the
fluctuations on a bulk brane or end-of-the-world brane as it moves through
the fifth-dimension. The fundamental papers on this subject can be found
in~\cite{stok,stokd,stoks,stok2,pn1,pn2}.

\subsection*{Acknowledgements}

Burt Ovrut is supported in part by the DOE under contract No.
DE-AC02-76-ER-03071.


\end{document}